\documentclass[preprint,5p,times,twocolumn,authoryear,sort]{elsarticle}
\usepackage{helvet}
\usepackage[T1]{fontenc}
\usepackage{amsmath,amsfonts,amssymb,mathrsfs,gensymb}
\usepackage{latexsym}
\usepackage{graphics}
\usepackage{graphicx}
\usepackage[pdffitwindow,%
	pdfstartpage=1,pdfstartview={XYZ 0 0 1},%
	pdfduplex=DuplexFlipLongEdge,pdfprintscaling=None%
	breaklinks,hyperindex,linktocpage,bookmarksnumbered,%
	colorlinks,allcolors=darkgreen]{hyperref}
\usepackage{epigraph}
\usepackage{epsf}
\usepackage{aniszewski}
\usepackage{color}
\usepackage[draft,textsize=small,color=green]{todonotes} %option 'disable' hides them I think
\usepackage[yyyymmdd,hhmmss]{datetime} %date is printed towards the bottom for version checking.
\usepackage{siunitx}
\usepackage{lipsum}

\title{Simulations of a Partially Submerged Cylinder Subjected to Current and Waves}

\author[coria]{Wojciech Aniszewski\corref{cor1}}
	\ead{aniszewski@pm.me}
\author[havre]{Valentin Ageorges}
\author[havre]{Gaële Perret}
\author[coria]{Vincent Moureau}

\ead{aniszewski@pm.me}

\address[coria]{CORIA, CNRS UMR6614, Normandie Universit\'e, UNIROUEN, INSA of Rouen, France}
\address[havre]{ Laboratoire Ondes et Milieux Complexes (LOMC) -- CNRS UMR 6294 -- Le Havre Normandy University, France}

\cortext[cor1]{Corresponding author at:  CORIA, 675 University Avenue - BP 12, 76801 St. Etienne du Rouvray, tel: (+33) 232 95 37 45, email: aniszewski@\{pm.me, coria.fr, dalembert.upmc.fr\}}

\journal{International Journal of Multiphase Flow}

\definecolor{mygreen}{rgb}{0.9,0.2,0.9}

\renewcommand\Re{{\rm Re}\,} %Reynolds
\newcommand\Fr{{\rm Fr}\,} %Froude
\begin{document}

\begin{abstract}
  This paper presents a new method allowing for a numerical study of the interaction between a submerged solid structure and a current with surface waves. The method involves a Navier-Stokes equation solver for a two-fluid formulation
  with the Accurate Conservative Level-set (ACLS) method used for interface tracking. The solver is generalized on unstructured grids, thus enabling a boundary-fitted implementation of the FSI (Fluid-Solid Interaction). A solid, vertical, partially
  submerged cylinder is examined as it moves through the flume. The latter is partially filled with flowing water; waves propagating on the liquid surface impact the cylinder. Simulation results include examination of the free surface topology,
  air entrapment and vorticity production in the cylinder wake, as well as the pressure distribution on the solid. Numerical results are validated against an experiment performed by the same authors. 
\end{abstract}

\maketitle

\section{Introduction}\label{intro}

%Bibliography notes on the subject with notes (will be transformed into continuous text later):

This paper presents a numerical study of the fluid-structure interaction in a realistic geometry, validated by an accompanying experimental investigation. The research presented here concerns a rigid structure in form of a partially submerged, vertical cylinder, which interacts with a two-phase flow -- namely a current with surface waves. In the laboratory setup, this takes place inside a $30$m-long flume. This type of flows occurs naturally in a number of coastal engineering problems, such as oil rigs subjected to oceanic waves~\citep{clearyrudman}, coastal installations subject to constant hydrodynamic loading, ship hydrodynamics or construction of bridge pillars. Depending on the scale of the problem, the studies in this broad field range from individual structures causing structure blocking~\citep{santo18} to bridge damage by hurricanes~\citep{padgett}. 

Studies of fluid-structure interactions, while spanning a wide range of spatio-temporal scales, dimensionless numbers and characteristic structure geometries, share certain challenges. Examples of these challenges are found in the   realistic representation of both the liquid(s) and the solids in the numerical simulation, or reliable and feasible measurement of the pressure and/or velocity field in the experiment -- especially for moving structures (see e.g.
~\citealp{lobovsky}). Thus, regardless of the shape or character of the structures or the nature of the flows, some of the problems will be universal. For numerical simulations, these include e.g. the robustness of the interface representation, especially near the solid.  For example, the work of \cite{fondelli} presents an interesting study of the hydrodynamic loading by means of both simulation and experiments. Namely, the authors are focusing on the dam-break flow in which the bulk liquid, once freed from the initial containment, impacts a solid, motionless block placed in its way on the container bottom (and equipped with pressure sensors).  The authors of \cite{fondelli} manage to reproduce the inertia-dominated stages of the flow, while the subsequent sloshing stages prove more problematic, most likely  due to aforementioned shortcomings in the interface tracking scheme -- which in their case is the Volume-of-Fluid method with the adaptive refinement of a structured mesh.

  \cite{Ai19} and \cite{xie20} present almost identical numerical simulations of a 3D singular wave interacting with a vertical cylinder (in both cases many more test cases are included for the presented codes). To represent the  interface, respectively the free-surface non-hydrostatic model and the simplified algebraic version of the VOF scheme (see e.g. \citealp{aniszewski2014caf}) are used. The channel depth there is only $0.04$m, an order of magnitude smaller than presented in this work. The authors of \cite{Ai19} impose the normal velocities of the incident periodic wave. Both \cite{Ai19} and \cite{xie20} include comparisons between simulations and experiment for the solitary wave interacting with the cylinder. Interestingly in the case of both papers, while the incident wave elevation and the hydrodynamic load are predicted well, there are discrepancies in the cylinder wake for both of the quantities. This is probably due to the deficiency of the simple interface descriptions used. This suggests that it would not be trivial to apply the simplistic interface treatment to the cases presented in our study, which include liquid overturning and/or air bubbles entrained in the cylinder wake. Other studies of  submerged structures of varying shapes can be found. For example, the authors of \cite{Arabi2019} study the interaction of a solitary wave with a rectangular impervious block. Their findings -- including a strong correlation between the structure's edges and the induced vortical structures in the wake --  concern also the  mechanism of vorticity production at the bottom of the partly submerged cylinder. A similar study to the one presented here is found in the work of \cite{carvaro}, wherein the authors use commercial software to study the wave-cylinder interactions at lower Reynolds numbers.

        Considering wave interactions with submerged structures, there exist a wide array of analytical \citep{tw97,davis74} and numerical (e.g. \citealp{brockl,han15}) works aiming at various structures - such as bridges, bridge decks and pillars, including flexible ones. For example, the work \cite{qu21} presents a numerical investigation into the wave interactions with a coastal structure.  While their focus is on the ''extreme'' waves (large amplitude oceanic waves originating from spontaneous focusing), the validation cases presented therein fit in the same flow regimes as presented in our work. Notably, the wave impact loads on a rectangular 2D cylinder are studied with results comparing reasonably to the experimental data. Moreover, the work of \cite{qu21} contains a numerical wave generation technique at the channel inlet similar to this work (see Sect.\ref{wavegensect}). However, the work of \cite{qu21} -- or comparable studies such as \cite{hu16} -- exhibits possible deficiencies in the interface description or simply lack of spatial resolution. This in turn could have made it harder for their authors to trace phenomena such as near-structure air entrapment or drag reduction in the wake. This problem is tackled in our work thanks to the unstructured, boundary-fitted meshes with variable resolution and a massively parallel code.

        The doctoral thesis one of these authors \citep{valentinPHD} and the companion publications \citep{valentinPRF,valentinFLUIDS} contain a study of the air entrapment in the flow around a partially submerged cylinder. The investigations there are both experimental and numerical using the YALES2 code~\citep{moureau2011} used in this work as well, for a wide range of Reynolds and Froude numbers. These works do not involve water waves: surface of the liquid is at rest before impacting the (moving) solid cylinder. The work identified two types of air entrapment (in a cavity along the cylinder wall and in the cylinder wake), and examined the critical velocities at which this phenomenon appears. Considering the cylinder drag, the work of \cite{valentinPRF} confirmed that the drag coefficient is generally smaller in multi- than in single-phase flow, an effect made even stronger by the air entrapment. The works in \cite{valentinPRF,valentinFLUIDS} are comparable to certain studies published using structured meshes (see e.g. \cite{elhanafi}) but avoids inherent numerical problems such as the solid modelling on rectangular grids, or limited resolution of the CFD simulations. The work presented here aims to continue the studies presented in \cite{valentinPRF,valentinFLUIDS}, by first refining the numerical methodology, and subsequently including the modelling of the incident waves in the fluid-structure interaction study. Unlike the previous works, here we focus on a single flow configuration -- corresponding to an intermediate value of the Reynolds number, intermediate wave amplitudes and low intensity of air entrapment in the cylinder wake -- and present a throughout comparison of the experimental and numerical approaches. 

\subsubsection*{Structure of the paper}

This paper is a part of a wider research project aimed at studying the liquid-structure interactions in varying configurations of hydrodynamic loading in single and multi-phase flows. Since our work involves both the experimental facility as well as the simulation framework, it is necessary to restrict the presentation here to its single aspect, for the sake of clarity. Thus, this paper is focused on presenting the proposed numerical methodology and the evidence of its validity. This dictates the following structure of the paper. In the Section \ref{y2sect} we introduce the numerical solver YALES2, describing the actual manner of solving the governing equations (\ref{codesect}), generation of the computational meshes (\ref{meshgensect}) the wave generators (\ref{wavegensect}) and the post-processing tools (\ref{postprosect}). Subsequently, we present a concise description of the experiment (Section \ref{expesect} along with references to more publications consecrated to it). In the Section \ref{simsect} we focus our attention on the results of the simulations, which we study in details including the macroscopic flow evolution (\ref{macrosect}) or the spectral analysis of certain signals (\ref{spectralsect}). The evolution of the pressure force acting on the cylinder is described in Section \ref{fxsect} for both the experiment and simulations. This is followed by a more detailed analysis of the interfacial waves (e.g. the reflection coefficient in \ref{reflectsect}), analysis of the flow statistics (\ref{statsect}), after which the paper concludes.

\section{Numerical methodology}\label{y2sect}

%\lipsum[2]

\subsection{Code Characteristics}\label{codesect}

YALES2\footnote{Yet Another Large-Eddy Simulation Solver} is a massively parallel, finite volume - based, structured and unstructured grid code, which began its existence as a combustion simulation tool \citep{moureau2011} for realistic, industrial geometries. It has since matured into a fully-fledged multi-physics code including nearly $40$ solvers. The code has been created at CORIA and is currently maintained by CORIA and a large global community of contributors from academia and industry\footnote{More information can be found on the YALES2 public website: \url{https://www.coria-cfd.fr/index.php/YALES2}.}. The code is written in Fortran, and employs -- within the F90 and F2008 language standards -- data structures and methods characteristic for object-oriented programming. Recent applications of YALES2 include the two-phase atomizing flow studies \citep{rj21}, new approaches to simulating  wind turbines \citep{benard18},  simulations of the Covid-19 viral transmissions \citep{dubief20} or proposing new numerical framework for high order approximation on finite volumes \citep{bernard20}.

To characterize the code briefly, we note that YALES2 is distinguished mainly by the internal data structures, created in the bottom-up fashion with massive parallelism in mind. The main grid object can link to a structured or unstructured mesh, with the latter being more optimized (and the former used mostly for testing and benchmarks in academic geometries). Thus, in realistic geometries, boundary fitted approach is used to model the domain boundaries. When working in parallel, most widespread approach to the CPU load distribution problem is Domain Decomposition (DD, see e.g. \citealp{chenzang17}), i.e. splitting the domain spatially and redistributing the grid nodes between the computing threads in a dynamical or static fashion. In YALES2, double-domain decomposition (DDD) is used, meaning not only thread (CPU) domains exist, but each of them has sub-domains called \textit{element groups}. An additional communication layer (internal communicator) exists in YALES2 on top of the external one (MPI), and the former permits information exchange between element groups and synchronization. This is justified on large distributed systems such as super-computers because while the cost of MPI communication remains similar in YALES2 to that using DD, its application of DDD greatly diminishes execution slowdowns due to memory cache misses and memory access on each of the computing cores. An additional benefit of DDD is that on each computing core, certain operations become possible without any MPI communications with neighbouring cores. An example of such an operation is generation of a coarse mesh e.g. for conditioning in the solution of the Poisson equation. To maintain optimal partitioning on all levels, the external METIS library \citep{karypis} is used. YALES2 is able to import simulation geometry in many formats (i.e. Ansys, HDF5, Gambit, GMesh) and is able to optimize and re-partition the grids after importing. In this paper, we utilize the said capability to an extent, importing a mesh and optimizing it using the external MMG3D library~\citep{mmg3d}. Considering the outputs of YALES2 simulations, HDF5 format is used along with in-solver optimizations which redistribute the input/output between threads to minimize the MPI communications.

\subsection{Governing Equations and Numerical Methods}\label{govern}

In this Section, we will present a brief description of the governing equations relevant to the work presented in this paper, followed directly by specifications of the numerical methods used to solve them. We solve the isothermal, incompressible two-phase fluid flow described by the Navier-Stokes equations in the so-called single-fluid formulation (see e.g. \citealp{delhaye}) :

\begin{equation}\label{nse}
  \frac{\partial \ub}{\partial t}+\nabla\cdot\ub\otimes\ub = \frac{1}{\rho}\left(\nabla\cdot\left(-p+\mu\mathbf{D}\right) +\sigma\kappa\nb\delta_s\right)+\mathbf{f_g}
\end{equation}

with the continuity equation

\begin{equation}\label{cont}
  \nabla \cdot \ub =0.
\end{equation}

In the above, $\ub,$ $p$ $\mu$ and $\rho$ stand, respectively, for liquid velocity, pressure, viscosity and density. The interface is denoted by $S,$ its normal vector as $\nb$ and its curvature as $\kappa;$ the delta distribution $\delta_S$ is centered on $S.$ In case of a discontinuity of any of the material properties between liquid phases, or nonzero surface tension coefficient $\sigma$,  the pressure may also be discontinuous on the interface $S$. This assumption is characteristic for the single-fluid formulation, although it is assumed that velocity is continuous on $S.$  The gravity (body) force is represented by $\mathbf{f_b}.$ The term $\mathbf{D}$ is the rate of deformation tensor \[\mathbf{D}=\frac{1}{2}\left(\nabla\ub+\nabla^T\ub\right).\] \textit{Note:} If needed, we denote material properties with ''l'' and ''g'' indices standing for liquid and gas; due to the single-fluid formulation these do note appear in (\ref{nse}) as will be explained below (see e.g. (\ref{rhodef}).

Restriction of the curvature to the interface $S$ is done using the distance function $\phi$, defined for each domain point $\xb$  as minimum of its distance from the interface:

\begin{equation}\label{phidef}
  \phi(\xb,t)\colon = \min (d(\xb,S)),
\end{equation}

where $d$ is a chosen metric. We thus note $\delta_S\colon = \delta_S(\phi)$. To denote restrictions to the interface we may also use the subscript $S,$ such as $\kappa_S\colon = \kappa\delta_S.$ When formulating the two-phase Navier-Stokes equations (\ref{nse}), the singularity of the interface $S$ is solved thanks to the  the pressure jump, denoted using the jump notation $\lbrack \cdot \rbrack_S = | \cdot_l-\cdot_g|.$ We thus define the pressure jump as

\begin{equation}\label{pjump}
  \lbrack p \rbrack_S = (p_l)_S - (p_g)_S = \sigma\kappa_S+2\lbrack \mu\rbrack_S\nb^T\cdot\nabla\ub\cdot\nb,
\end{equation}

where $\lbrack \mu \rbrack$ is the viscosity jump if applicable. In all the above, the normal vector $\nb$ can also be defined using the distance function as

\begin{equation}\label{ndef}
  \nb = \frac{\nabla\phi}{|\nabla\phi|}.
\end{equation}

In the traditional level-set method formulation, the advection of the distance function $\phi$ defined in (\ref{phidef}) can be realized by a direct solution of its advection equation \[\frac{D\phi}{Dt}=0\], that is zeroing of the material derivative. This is made possible by the fact that $\phi$ is continuous and smooth at least in the interface vicinity. However, numerous works have indicated that the method is lacking in terms of mass conservation of the traced liquid, especially in comparison to the Volume-of-Fluid (VOF) type schemes (such comparisons can be found e.g. in \citealp{aniszewski2014caf}). Therefore, this work employs the Accurate Conservative Level-Set (ACLS) method implemented in the YALES2 framework as described in \cite{rj21}. In this method, a new scalar field $\psi$ is defined, using the hyperbolic tangent of the distance function:

\begin{equation}\label{psidef}
\psi(\xb,t) = \frac{1}{2}\left( 1+\tanh\left( \frac{\phi(\xb,t)}{2\epsilon} \right)\right),
\end{equation}

where $\epsilon$ sets the profile thickness by modifying the steepness of $\psi$ near the interface. By convention, $S$ is now defined by \begin{equation}\label{convent}S(t)=\left\{\xb\in\mathbf{R}^3\colon\psi(\xb,t)=\frac{1}{2}\right\},\end{equation} which is equivalent to setting $\phi=0$ in most cases\footnote{Equation (\ref{convent}) allows us to count ACLS among the sharp interface methods, although it is interesting to note that the definition of $S$ can be broadened by reusing the VOF convention of considering as interfacial all grid cells for which $0<\psi(\xb,t)<1$; this results in a diffuse interface method.}. Due to similarity to the $C$ function used in the VOF methods, we may occasionally refer to $\psi$ as the volume fraction function although the definition (\ref{convent}) is not identical to VOF volume fraction functions. However, $\psi$ assumes the value $0$ in one phase and $1$ in the other, as $C$ does, which justifies this naming.

Having defined $\psi,$ we can advect $S$ by a solving the equation:

\begin{equation}\label{psiadv}
  \frac{\partial\psi}{\partial t}+\nabla\cdot\left(\psi\ub\right)=0.
\end{equation}

In YALES2, the solution of (\ref{psiadv}) is achieved via 4-th order spatial discretization and the TFV4A time advancement scheme \citep{rjr42}. As is the case with the LS method, ACLS requires re-initialization of the hyperbolic tangent function, which in our work uses the approach of \cite{rjr27}. Like the normal vector $\nb$, the curvature $\kappa$ is calculated directly from $\phi$  (see \citealp{rj21} for the implementation details).

Using $\psi,$ we are able to define the material properties for use in (\ref{nse}). We utilize the definitions from the Volume-of-Fluid type methods (e.g. \citealp{Xie17}):

\begin{equation}\label{mudef}
\mu(\xb,t)=\mu_g+\left(\mu_l-\mu_g\right)\psi(\xb,t),  
\end{equation}

where $\mu_l$ and $\mu_g$ are constant viscosities of liquid and gas, respectively. With this smooth $\mu$ definition, intermediary velocity values will be encountered in the near-interface grid nodes. Unlike (\ref{mudef}), the definition of $\rho$ is a sharp one and utilizes the Heaviside function $H(x)$:

\begin{equation}\label{rhodef}
\rho(\xb,t)=\rho_g+(\rho_l-\rho_g)H\left(\psi(\xb,t)-\frac{1}{2}\right).
\end{equation}

This ensures that node-to-node density jumps will be coupled with the phase definitions provided by (\ref{psidef}), thus limiting the risk of non-physical momentum fluxes \citep{geo2015}.

\begin{center}$\star$\end{center}

Considering the solutions of equations (\ref{nse})-(\ref{rhodef}), we first note that YALES2 is a projection method \citep{tsz} based solver. This means that the Poisson equation for pressure will be solved; in this work this happens via the Deflated Preconditioned Conjugate Gradient (DPCG) method \citep{rjr39} used to solve the resulting linear system. Inclusion of the pressure jump (\ref{pjump}) in the above solution is done via the Ghost Fluid Method (GFM) which in our case is an unstructured grid formulation of \cite{fedkiw-gfm}.

Similar to the coupled Level-set Volume-of-Fluid (CLSVOF) methods \citep{aniszewski2014caf} the coexistence of the hyperbolic tangent function $\psi$ and the distance function $\phi$ in ACLS leaves room for some choices in the implementation. The functions can be advected independently and coupled (the CLSVOF analog is presented in \citealp{aniszewskiJCP}) or they can be re-initialized periodically (or recreated) one from another (see \citealp{menard}). In the YALES2 implementation of ACLS, the $\psi$ function is advected by solving (\ref{psiadv}) and $\phi$ is actually recreated from $\psi$ at each solver time-step. Then, the distance function $\phi$ can be used to calculate $\nb$ and $\kappa$. While the details of the $\phi$ calculation algorithm can be found in \cite{rj21}, we summarize it briefly here. First, YALES2 uses a \textit{narrow band} of grid elements on both size of $S$ to define $\phi$ only there, which reduces the CPU cost\footnote{The thickness of this narrow band is a user-controlled parameter}. Second, the reconstruction of $\phi$ from $\psi$ is done using a geometrical Fast Marching Method in the framework of the GPMM (Geometric Projection Marker Method) of \cite{rjr35}. It involves the full 3D triangulation of the interface on tetrahedral grids, which is followed by projecting interface markers onto the triangulated surface. The line along which each marker was projected is the first approximation of the distance function gradient. Hence $\phi$ can be reconstructed from this information by propagating the triangulated surface via the Fast Marching Method inside the narrow band. This algorithm is optimized for massively parallel calculations as described in \cite{rj21}.

In closing of this Section, we denote that the distance function $\phi(\xb,t)$ is essential for this work as it can be used to directly obtain the distance between $S$ and selected grid nodes. For a point $P(x_p,y_p,0)$ where $x_p$ and $y_p$ are coordinates on the initially flat liquid interface, the elevation $\eta(t)|_P$ is approximately equal to $\phi(P,t)$ which we use to generate the signal of the point probes (see Sect. \ref{macrosect}). In general $\eta$ and $\phi$ can only be exactly equal if the local curvature radius $R_p$ at $P$ will be greater than the distance between $P$ and $S$: \[\phi(P,t)=\eta(t)|_P \Leftrightarrow d(S,P) \ll R_p.\] While this condition is fulfilled for the probes used in this work, it has to be noted that even the failure to observe it results in errors that are of the order of the grid size near $S$, which is one to two orders of magnitude lower than the $\eta$ signals measured.

\subsection{Wave Generation}\label{wavegensect}
%(in which we actually may write something interesting)
%\lipsum[4]

The wave generation technique used in this paper is based on the work \citep{rapp90} which introduced the multi-modal wave packet description using the velocity potential $\Phi$. Unlike the solutions based on the piston model \citep{chenzang17} or momentum sources \citep{ha13}, this approach is well suited for solitary or uni-modal waves.  First, the explicit expression for the liquid elevation writes as

  \begin{equation}\label{rapp1}
    \eta(x=0,t) = \sum_{i-1}^N a_i \cos [\Omega_i(t-t_f)-k_i(x-x_f)],
  \end{equation}

where $N$ is the number of modes. Formula (\ref{rapp1}) is not potential-dependent, and can easily be implemented as a Dirichlet-type inlet condition for the hyperbolic tangent function (\ref{psidef})\footnote{As a matter of fact, it is possible to define naive wave generators using only (\ref{rapp1}) and a flat velocity profile, however these simple techniques do not perform well in realistic loading studies.}.  This is followed by velocity components defined as derivatives of $\Phi$:

  \begin{equation}\label{rapp2}
    u(x=0,t) = u_0+\frac{\partial\Phi}{\partial x} \quad w(x=0,t) = \frac{\partial \Phi}{\partial z } ,
  \end{equation}

while $v(x=0,t)=0.$ The system is closed by defining the potential

  \begin{equation}\label{rapp3}
    \Phi (x,z) = \sum_{i=1}^{N} \frac{a_i r(t)\Omega_i}{k_i} \frac{\cosh ( k_n (z+d))}{\cosh k_n d}  \sin [\Omega_i (t-t_f)-k_i (x-x_f)].
  \end{equation}

In the above equations, $a_i$ $k_i$ and $\Omega_i$ are respectively the amplitude, wavenumber and the angular frequency of each mode. Water depth is represented by $d.$ The $t_f$ and $x_f$ characterize the focusing location of the wave packet. The above equations are imposed as the time-dependent Dirichlet boundary condition at $x=0.$ Additionally, the numerical ramping function $r(t)$ is used in (\ref{rapp3}), defined as \[ r(t)=2\arctan(8t)/\pi\] which ensures stability of the generator in the initial stage of the flow (note $r(t)>0.96$ for $t>2$).
  
The system (\ref{rapp1})-(\ref{rapp3}) has been proposed to produce multi-modal packets with a set focusing location e.g. to generate breaking waves. However, as described in Sect. \ref{expesect}, the simulated configuration consists of a series of  identical waves which interact with the cylinder cart. Therefore, we have decided on setting $N=1$ in (\ref{rapp2}), and setting $a_i$ and $k_i$ to the values measured experimentally. Additionally, we include  $u_0$ in (\ref{rapp2}) corresponding to the cart velocity (for the \textit{h5c3} configuration, $u=0.837$m/s). For $u_0\ne 0,$ using the ramping function $r(t)$ ensures the relaxation effect similar to that of a fixed, near-inlet  relaxation zone used e.g. by \cite{qu21}, except in the reference frame of the moving cylinder. It is easy to check that assuming $N=1,$ equation (\ref{rapp2}) reduces to

\begin{eqnarray}
  u(x,t) & = & u_0 - r(t)a_1\Omega_1\cos(kx-t\Omega_1-p_s)\cdot\\ \notag
         &   & \cdot\frac{\cosh(k_1z+dk_1)}{\cosh(dk_1)}\mbox{ } \label{g2u}
\end{eqnarray}

and

\begin{eqnarray}
  w(x,t) & = & -r(t)a_1\Omega_1\sin(k_1x-t\Omega_1-p_s)\cdot\\ \notag
         &   & \cdot\frac{\sinh(k_1z+dk_1)}{\cosh(dk_1)},\mbox{ }\label{g2w}
\end{eqnarray}

where $p_s$ is the phase shift term equal to $\Omega_1t_f+k_1x_f.$ Since the wave generator matches the experimental parameters as explained in Sect. \ref{expesect}, the wavenumbers and frequencies visible in (\ref{g2u}) and (\ref{g2w}) must obey the dispersion relation (\ref{dispe2}) for the waves observed in the flume. With (\ref{dispe2}), the above system is closed and can be implemented directly for any value of $N.$

%%   \red{UNFINISHED! RAW!}
  
%% \red{TODO:
%%   \begin{itemize}
%%   \item re-read and check all symbols
%%   \item Dispersion relation
%%   \item Sponges as options. Due to the fact that the $C_r$ study for the $1$m-long sponge gives strange results, I can use this fact to justify the fact that sponges are not present in all calculations.
%%   \item (Can mention Y2 inlet-outlet coupling?)
%%   \end{itemize}
%% }

\subsection{Post-processing}\label{postprosect}

The YALES2 solver allows for the simulation data acquisition in multiple formats (ASCII and binary), thus we will restrict this description solely to the means used in this work.

The acquisition of the simulation scalar and vector fields is realized in parallel using the HDF format. The solver uses HDF5 both for structured and unstructured data; automatic balancing is used to distribute the I/O operations between the used threads. However, the data chunks saved onto the disks need not reflect the structure of the MPI threads. So for example, it is perfectly possible to use 1024 processors which dump the data to 28 HDF5 chunks. Further processing of the data is possible using widely known visualization suites such as Paraview \citep{paraview}. \textit{A priori} all the scalar and vector fields are periodically dumped during the simulation.

To offer a more fine-grained insight into the flow physics, we employ probes. These are either of the predefined sampling points, lines or planes, over which the acquisition is taking place at the specified time intervals. This allows extraction of characteristics such as profiles or temporal evolution of given scalar/vector fields (see e.g. Figure \ref{probemix}). In the simulations presented here, we have opted for the point probes P1-P9 which we use to extract the interface elevation $\eta$, local velocity magnitudes or pressure; details of the placement of the probes are given in Table \ref{probetab} (see also Section \ref{macrosect}). Where possible, the YALES2 probes are meant to correspond exactly or closely to the placement of the resistive probes in the experimental setup, as explained below. (When mentioning the experiment  probes, we will denote them with the ''exp'' subscript, such as P$3_{exp}$. )

Other information means are offered by several kinds of statistical quantities. Starting from the lowest order statistics: the minimum, average and maximum values of chosen scalar fields are stored as they evolve in time. Typical uses here involve quantities such as $\max(\ub(t))$ or $(\int_V E_K dV)/V$. Moreover, YALES2 computes the mean (expectation) of $\ub$ understood as a random variable. Included are also higher moments such as standard deviation (the root mean square, or the r.m.s.). This can be performed also for other scalar fields such as the interface tracking functions, and will be discussed in Section \ref{statsect}. Finally, the solver computes statistics using the phase averaging procedure, as will be specified in Section \ref{statsect}. Some statistical values, as well as quantities describing the code performance, can be acquired from the solver logs.

For the majority of the data presented in this work, BASH and Gnuplot scripts are  used to automatize the presentation of the results, especially concerning the ASCII data (probe signals, time traces). This post-processing involves also the Fourier transforms (see Section \ref{spectralsect}) employing the FFTW3 \citep{FFTW05}. 

\subsection{Mesh Generation}\label{meshgensect}

In this subsection, we will briefly describe the procedure of generating the computational grid mesh(es) for the simulations presented in this work. Additionally, a description of the actual used meshes will be given. The grid generation procedure involves a specially constructed YALES2 test case whose purpose is the import of a ASCII STL file describing the solid immersed in the computational domain with arbitrary physical dimensions. In the procedure, an ASCII file describing the solid is imported into YALES2 and placed in a domain of arbitrary shape. Subsequently the STL geometry specified by the file is re-meshed using tetrahedrons at a specified minimum spatial resolution. The user controls parameters such as element skewness or the maximum mesh size gradient $\max|\nabla(\Delta x)|$. In an intermediary stage we specify additional refinement zones (.e.g. for the air-liquid interface), or coarse grid zones (''sponges'') which may be used e.g. for numerical damping of kinetic energy and testing purposes. Certain details of the used computational meshes are given in Table \ref{meshtab}.

\begin{table*}
  \begin{center}    
    \begin{tabular}{l l l l l}
      Name &  No. Elem. ($\cdot 10^6$)  & $\min(\Delta x)$ & $x$-span & Commentary \\ \hline
      G0     &  $21.8$                    & $5$mm            & $\pi$m    & Tests, shorter cylinder, less refinement \\
      G01     &  $39$                      & $5$mm            & $\pi$m    & Main production \\
      G02     &  $71.8$                    & $4$mm            & $\pi$m    & Convergence study \\
      G03     &  $160$                     & $3$mm            & $\pi$m    & Convergence study \\
      G41   &  $35.8$                    & $5$mm            & $\pi$m    & $C_r$ study, no cylinder \\
      G42   &  $39.7$                    & $5$mm            & $4.141$m  & $C_r$ study, longer, sponge layer \\ \hline
    \end{tabular}
    \caption{Characterization of the computational grids (meshes) used in the simulations. For all meshes, their physical dimensions are  $x\mbox{-span} \times 1 \times 0.6$m. All grids are tetrahedral. }
    \label{meshtab}
  \end{center}
\end{table*}
%Notes and TODO  for the table:
%1. G0 had the wrong cylinder length - which is why 21m
%2. G4.2 has had a high skewness up to 0.9, 39.74M elements and was created on 2021-10-03 (Myria).

\begin{figure}[ht]
  \centering
  \includegraphics[width=0.8\columnwidth]{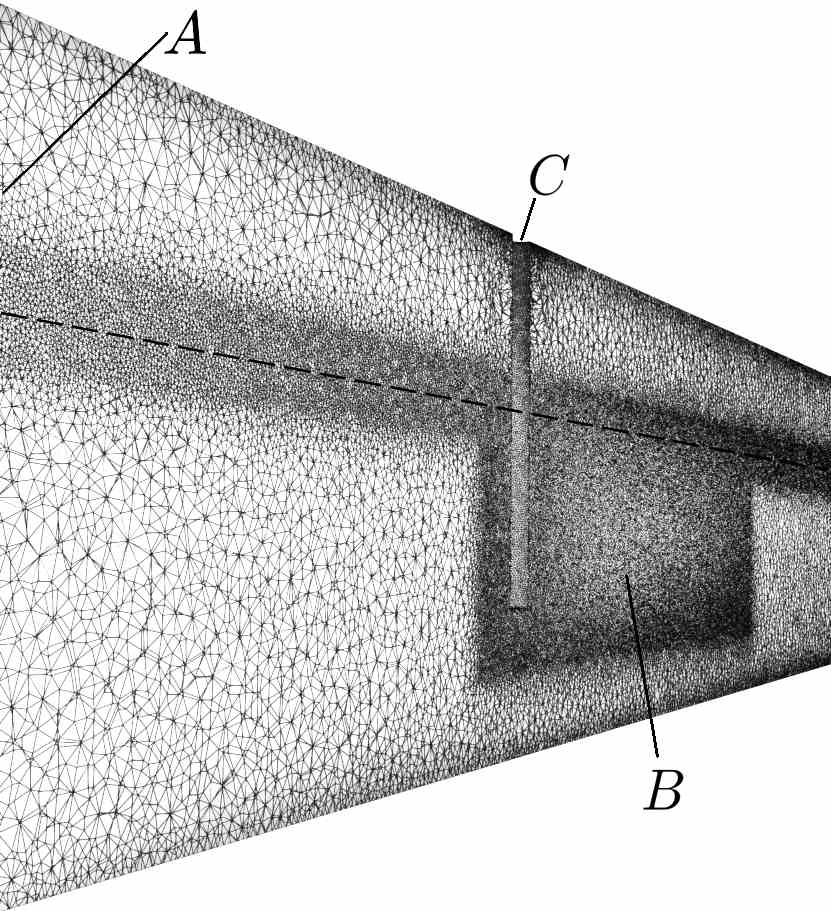}
  \caption{The grid G01 as specified in Table \ref{meshtab}. The grid is visible as a cut-volume along the $xz$ plane. A - inlet, B - the fine mesh box containing the cylinder wake area, C - the cylinder visible as a boundary fitted opening in the mesh. The liquid/air interface location is presented with a dashed black line. }\label{g1fig}
\end{figure}

We denote that YALES2 employs the AMR (Adaptive Mesh Refinement) on run-time, however in this work we use AMR only in the limited sense for grid generation at the beginning of the simulations, the grid is then kept static. Specifically, after the STL file describing the cylinder is imported and the fine/coarse mesh zones prescribed, a preliminary simulation is ran which uses AMR over several dozens of solver iterations until the mesh converges to specifications. In this preliminary simulation, the MMG3D library \citep{mmg3d} is used to perform the grid adaptation (see also \citealp{benard16}).  Subsequently, thus obtained mesh is moved into a separate, production simulation. The rationale for this procedure is that in our particular flow case it is possible to predict e.g. the interface location in the quasi-stationary flow stage. Even once the flow is fully developed and quasi-stationary, it involves waves of known amplitude (approximately $2a$) thus the interface resides in a zone of given thickness ($z$-span). We additionally refine the  vicinity of the solid cylinder and a box-shaped volume in the cylinder wake, as shown in the Figure \ref{g1fig} which showcases the mesh G0 as specified in Table \ref{meshtab}. One notes by inspecting Figure \ref{g1fig} that any possible air entrapment in the cylinder wake will take place in the refined zone as well. This static-AMR approach circumvents the need of a fully fledged AMR on an unstructured mesh thanks to the predictability of the interface residence areas -- this simplifies the simulation procedure and improves the repeatability of results. Note that simulations of several other two-phase flow cases  have been published using YALES2 \citep{rj21} which operate the AMR in a fully dynamic manner.

%brief description and rationale for the \meshtab grids
Of the grids described in Table \ref{meshtab}, the grid G01 has been used as a production grid providing majority of the results presented in this paper, except when indicated. The G0 grid was used only in the testing stage, and was, like G01,  characterized by the $3.14$m domain length, however a different cylinder shape was used than in G01, also the refinement areas in the cylinder wake were significantly smaller, limiting the G0 grid to $21$ million elements. The grids G02 and G03, carrying respectively $71.8$ and $160$ million elements, are designed for the convergence study. Their respective minimal spatial resolutions are $4$mm and $3$mm, otherwise they have identical specifications to the G01 grid. Finally, two grids have been prepared as a part of the investigations of the wave reflection phenomena (see Sect. \ref{reflectsect}) and calculations of the reflection coefficient $C_r.$ The first of the grids, G41 contains the channel geometry without the immersed solid cylinder, which is a rudimentary method to ascertain the degree to which the latter contributes to the (measured) wave reflections in the simulations. The resolution is kept equal to the main production grid G01 and the number of elements is $35.8\cdot 10^6.$
 The other conceivable source of the numerically-induced reflections is the outflow boundary condition, a possibility we investigate using the last of the grids presented in Table \ref{meshtab}: G42. Compared to the main grid G01, the G42  has been elongated by an additional $1$m downstream. Moreover, the additional domain volume thus obtained acts as a numerical sponge layer, since it is left unrefined -- which explains a relatively low element count  of G42 ($39.7$ million) compared to G01, with only $7\cdot 10^5$ elements added. Adding sponge layers is a rudimentary method of (numerically) absorbing the kinetic energy, and has been well tested in wave hydrodynamics simulations \citep{ha13, israeli81} including the wave-cylinder interactions \citep{Ai19}. In the case of the G42 grid, the unrefined elements in the sponge area have sizes close to $\max(\Delta x)=0.05$m; no additional modifications of the velocity field or the interface are used.

\section{Experiment Specification}\label{expesect}

In this Section, we present a brief description of the experimental setup. The experiments have been performed using a cylinder of $D=5$ cm made of polymethyl methacrylate (PMMA) and clogged at its free-end. The cylinder is mounted on a traction cart moving through a wave flume of 34 m in length, $W=90$ cm in width and 120 cm in height filled with $d =0.4$ m of tap water. The cylinder is partially immersed with a depth $h_d = 0.23$ m in tap water, and moves along the flume with a velocity, $u_0=0.8375$ m/s.

\begin{figure}[ht]
  \centering
  \includegraphics[width=\columnwidth]{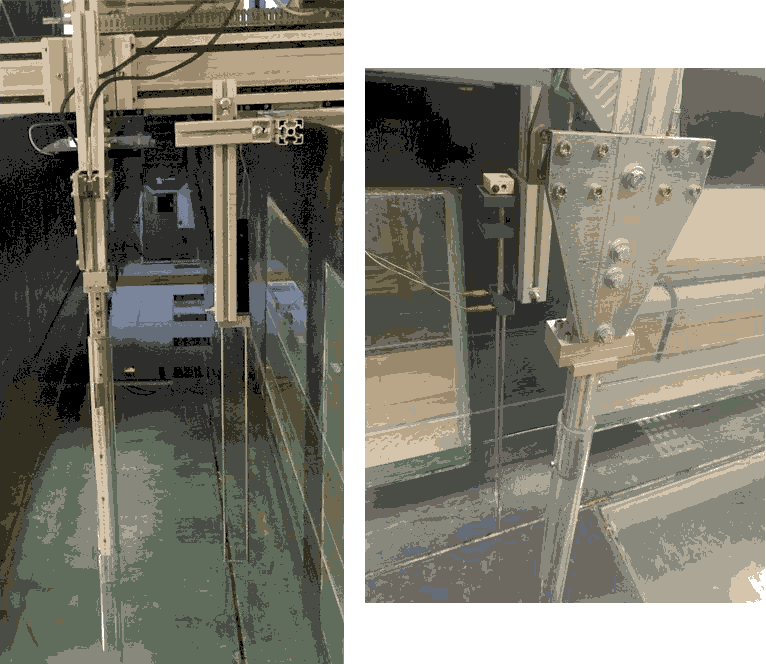}
  \caption{Configuration of the experiment with the cylinder of diameter $5 cm$ and the resistive probe allowing the use of the point method of determination of the adimensional coefficients.}\label{manip}
\end{figure}

Previously, a set of experiments, where cylinders have been translated at different constant velocities through water initially at rest has been achieved with the similar setup. For those experiments, the focus had been brought on the drag and air-entrapment phenomena for different cylinder diameters \citep{valentinPRF}.
In the current study, the waves are generated using a piston type paddle. The incident/reflected wave amplitudes and associated reflection coefficient $C_r$ are estimated using the method of Mansard and Funke \citep{mansard1980} with two resistive probes placed upstream the cylinder (see also Figure \ref{manip}).

In the frame of reference of the laboratory, the incoming waves have the period $T=2$ s, the amplitude (crest to through) $2a =0.048$ m and a maximum celerity at free-surface $u_{FS} \simeq 0.127$ m/s (calculated from Stokes theory at first order).  The characteristics of the incoming waves are determined in the frame of reference of the laboratory with probes placed at P$3_{exp}(-1.5,-0.3,0)$ and P$2_{exp} (-0.7,-0.3,0)$ of the fixed cylinder and a sampling frequency of $100$Hz. \textit{Note:} these probe locations correspond  to the probe points specified in the  Table \ref{probetab} for the simulations. 

In Figure \ref{manip}, we present the configuration for the experimental study of the problem. For the reminder of this description, we will, when specified, use the frame of reference attached to the cylinder which is able to move with the velocity of current $u_0$. 
A resistive probe is visible in both pictures next to the cylinder at P1 position (see also Table \ref{probetab}).
At the upper part of the Figure \ref{manip}, a piezoelectric sensor (Kistler 9327C), with a sensitivity of -7.8 pC/N  allows the measurements of the drag force $F_x$, acting on the whole cylinder. 
The sensor is connected to a charge amplifier (Kistler 5015A) equipped with a low pass filter of 30 Hz, itself linked to a LabView acquisition card.
Signals are filtered using a moving average on 20 points.
The sensor and the resistive probe are synchronised and have a frequency acquisition of 100 Hz.\\

The relation between frequency and wave number is given by the dispersion relation for capillary-gravity waves:

\begin{equation}\label{dispe}
  \Omega^2 = gk\tanh(kh),
 \end{equation}

for $k<<k_c$ with $k=\frac{2\pi}{\lambda}$ the wave number, $k_c \simeq$ 367 m$^{-1}$, the capillary wave number, $\Omega=\frac{2\pi}{T}$ the pulsation, $g$ the gravity. In the frame of reference of the laboratory, the wave number is  equal to $k = 1.7005$m$^{-1}$. 

In the reference frame attached to the cylinder adding a current by moving the cylinder at constant velocity $u_0$ leads to the modification of the dispersion relation (\ref{dispe}). This Doppler effect modifies it via a frequency shift of $ku_0,$ namely

\begin{equation}\label{dispe2}
  \left(\Omega + ku_0\right)^2 = gk\tanh(kh)
\end{equation}

leading to a period of 1.43 s in the cylinder frame of reference. The same dispersion relation is implemented in the numerical wave generator, so that the potential equation (\ref{rapp2}) can be closed.

\section{Simulation Results}\label{simsect}
%\lipsum[6]

\subsection{Specification of the Test Cases}\label{simspec}

The simulation utilizes a domain whose physical dimensions are meant to closely resemble the actual experimental structure found at the LOMC. YALES2 makes it conceivable even to include the entire experimental geometry: the $34$m-long tunnel, a moving measurement cart and a wave-generating beater -- since the solver includes provisions for moving solids with attached grids, adaptive mesh refinement and is generally formulated along the LES methodology \citep{pope} so that arbitrary precision levels can be chosen throughout the domain. However, the associated CPU and GPU cost of such a holistic approach would be prohibitive and unjustified, as there is hardly any need to e.g. study the far-wake phenomena in the presented case. Thus, these  authors have opted for a more restricted study, which on the other hand allows for higher spatial resolutions, the DNS-type methodology and inclusion of all transient effects.

More specifically, the simulation domain is built using the frame of reference of the cylinder: it remains at the point $(0,0,0)$ while the liquid flows around it, with an inlet and outlet at both domain extremities. The cylinder is placed along the $z$ axis, perpendicular to liquid interface at rest  which stretches in the $xy$ plane. The flow is from $x-$ to $x+$ while the $y-$ and $y+$ domain walls are referred to as ''front'' and ''back''. The sizes of the domain are chosen as $3.14 \times 1 \times 0.6$m which corresponds to a $3.14$m-long section of the experimental flume. The inlet is placed at $x=-1$m, thus situating the outlet at $x=2.14$m. Water depth (zero-level, i.e. disregarding the wave crest/through) is $d=0.4$m. Unless noted otherwise, the results presented here correspond to the \textit{h5c3} experiment configuration, which is a naming convention signifying the cylinder diameter $D=0.05$m, presence of water current with the mean inlet velocity $u_0=0.8375$m/s, and the wave amplitude (crest to through) $2a=0.048$m. For the \textit{h5c3} configuration, we denote the values of the Keulegan-Carpenter number of \[K_c\colon = \frac{T\left(\ub_{FS} \right)}{D} \approx 3.5, \] while the Reynolds number $\Re=41875,$ resulting in the ratio $\Re/K_c\approx 11980.$

It is essential to note that the solid cylinder is, as in the experimental setup, immersed only partially to the depth $h_d=0.23$m, thus leaving $17$cm of spacing between itself and the channel (flume) bottom. Since our goal is to simulate the moving measurement cart in its reference plane the boundary conditions for the \textit{front} and \textit{back} walls are translating wall (translation velocity set to $u_0$). No-slip walls are chosen for the \textit{bottom} and \textit{top} walls as well as the cylinder surface. A free outflow is configured for the $x+$ domain outlet, while the inlet condition is coupled with the wave generator, as explained in the Section \ref{wavegensect}. A good overall look at the domain shape is provided by Figure \ref{t3sc}, in which the flow direction is from top to bottom, and the cylinder (not rendered) is marked by an opening in the liquid interface, the later coloured by $|u|.$

%new piece 2021-10-18 on the CPU cost associated.

Considering the associated CPU cost of the presented simulations, it has been controlled mostly by the CPU architectures/generations of choice. Using the G01 grid ($39$ million elements) as reference , the cost ranged from $50\cdot 10^3$ CPUh when using Intel CPUs (Broadwell x86) down to $25\cdot 10^3$ CPUh using AMD (EPYC2 x86 Rome) cores -- for a simulation spanning approx. $20$s of physical time. Breaking down the algorithm cost per iteration results in a even split between the LS advancement and redistancing ($\approx 45\%$ per iteration) and the Poisson equation solver ($\approx 43 \%$). Looking in turn at the \textit{cumulative} (total) cost of the presented YALES2 simulations, the i/o operations have a significant share in the CPU cost at approximately $30\%$ (this however is strongly hardware-dependent).

\subsection{Macroscopic Flow Evolution}\label{macrosect}

%\lipsum[8]

\begin{figure}[ht!]
\centering
\includegraphics[width=.99\columnwidth]{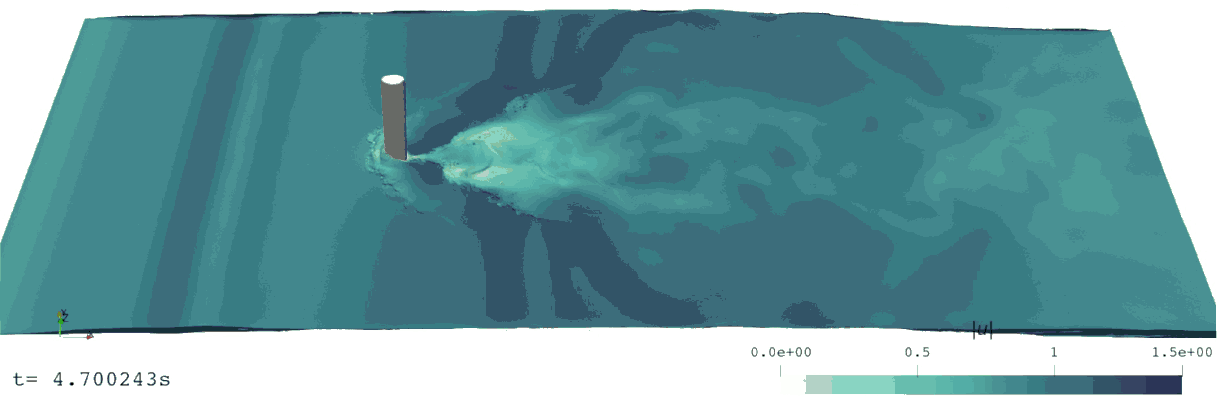}
\caption{Macroscopic flow geometry for the YALES2 simulation of cylinder-wave interaction. }\label{32a}
\end{figure}

The macroscopic evolution of the interface for an example simulation is shown in Figure \ref{32a}.  The surface is coloured according to the magnitude of the velocity. The vertical cylinder is not rendered visible in Fig. \ref{32a}, except as an opening in the interface. The wake is however clearly visible behind it. The cylinder position of the structure is  $1$m from the inlet. In Figure \ref{32a}, we are observing an incoming wave upstream, the ''v-shape'' liquid folds at the borders of the cavity, as well as the traces of the vortical structures on the liquid surfaces downstream. More accurate readings of the interface elevation $\eta(\xb,t)$ are given below using the probes. At this stage, we only denote that the result presented in Fig. \ref{32a}, obtained using the $5$mm grid G01  closely resembles the macroscopic flow images obtained experimentally at this Reynolds and Froude numbers \citep{valentinPHD}. The level of detail e.g. considering the interface geometry is far superior to certain previous works -- for example  \citep{han15} where the authors have used even finer meshes near the immersed structure\footnote{The cited work concerned lower Reynolds numbers in somewhat different context, but the obtained geometrical information concerning the interface, as well as the details of the $\ub$ field, are much less detailed.} and compares well to works such as \cite{koo} whose authors have investigated individual vortical structures around the immersed structure at high Reynolds numbers.  The same grid had been used for the result present in Figure \ref{32b} below.

\begin{figure}[ht!]
\centering
\includegraphics[width=.99\columnwidth]{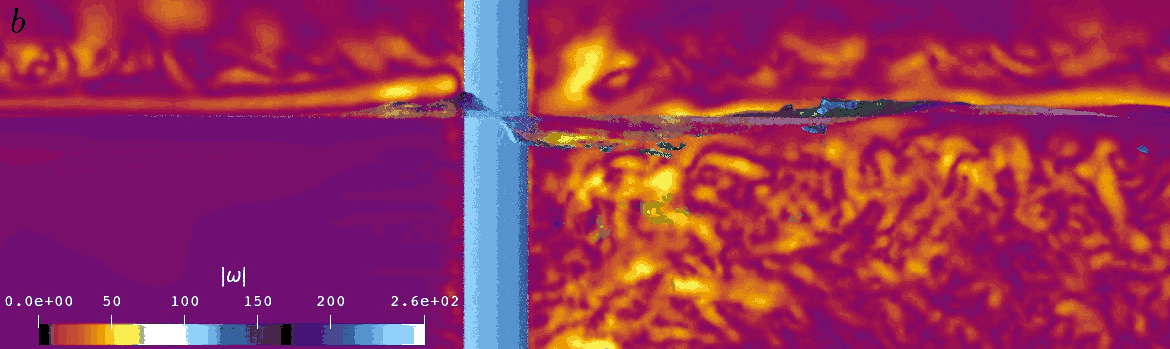}
\caption{ The geometry of the near-cylinder interface for the same time ($t\approx 24.8$s) as in Fig. \ref{32a}. }\label{32b}
\end{figure}

Figure \ref{32b} shows the same moment in the example simulation as in Figure \ref{32a}, except that this time we focus on the vicinity of the cylinder. The cylinder is visible in blue, the interface is made semi-transparent, showcasing the liquid elevation at the cylinder's upstream side and a slight elevation drop downstream. The plane visible in the background is coloured using the vorticity norm or $|\omega|$, showcasing rich vortical structures in the wake. The authors of  \cite{gonclaves} whose work concerns an experimental PIV measurements for a case without waves\footnote{ However, it features a flow characterized by $\Re\approx 43000$ which is close to our configuration, even if the $h_d/D$ ratio is lower ($h_D/D\approx 2$ compared to $h_D/D\approx 4.6$ here.} have established that the  vorticity is produced principally in two areas: one below the cylinder's bottom, and another present as a circulation zone directly adhering the cylinder downstream.  The action of the latter circulation zone is well visible in Fig. \ref{32b} due to high levels of vorticity.

Concerning $\omega,$ the authors of \cite{valentinPRF} have established that the air entrapment in the vicinity of the partially submerged structure acts as to mitigate the vorticity generation in its wake; however their findings were for $\Re>100000,$ while the case discussed here is barely in the intermittent entrapment regime meaning that the bubbles are either not at all, or rarely visible. Nonetheless, unlike their work, we have focused on the wavy flow, which itself is a mechanism modifying $\omega$ as will be discussed below in context of Figure \ref{vortfx}. As said the macroscopic evolution of the flow resembles closely the experimental results. The wake of the cylinder evolves during the passage of the waves, we also observe the intermittent entrapment of the air reappearing with the frequency $1/T$; in Figure \ref{32b} this is visible in the form of small, elongated bubbles colored blue near the cylinder. 

\begin{figure}[ht!]
\centering
\includegraphics[width=.7\columnwidth]{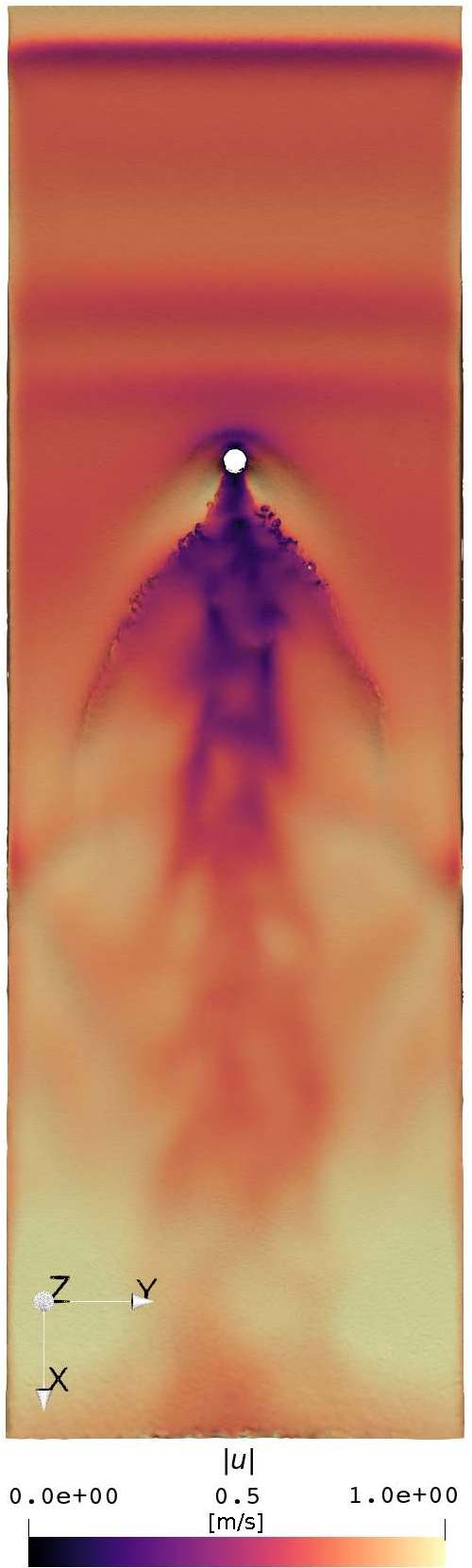}
\caption{Macroscopic flow geometry for the YALES2 simulation of cylinder-wave interaction: top view, coloured by $|u|$. The flow direction is from top to bottom.}\label{t3sc}
\end{figure}

Figure \ref{t3sc} presents the top-camera position view of the entire interface geometry in the $3.14$m - long flume at the instant of $t\approx 3$s. Colouring is done using the velocity magnitude. An incoming, generated wave is clearly visible close to the inlet, while the cylinder wake is well visible below it. The liquid folds forms a neck behind the structure (the ''v-shape'') whose angle can be measured both for the experimental as numerical data (see Figure \ref{halfangels}). This instantaneous result presents also slight interface deformations (visible at the outflow edge). This numerical phenomenon occurs occasionally in YALES2 due to strong $\Delta x$ in the cells passed by the interface (we noted it only in the G01 grid). This can be avoided using fully dynamic AMR or finer meshes in general (such as G02 and G03), however even in our meshing strategy these kind of errors appeared only penultimate cell layer next to the outflow, and the error has no means to propagate upstream. Further looks at the interface geometry will be provided below (e.g. Figures \ref{phiavg3} and \ref{render}).

\begin{figure*}[ht]
  \centering
  \includegraphics[width=.7\textwidth]{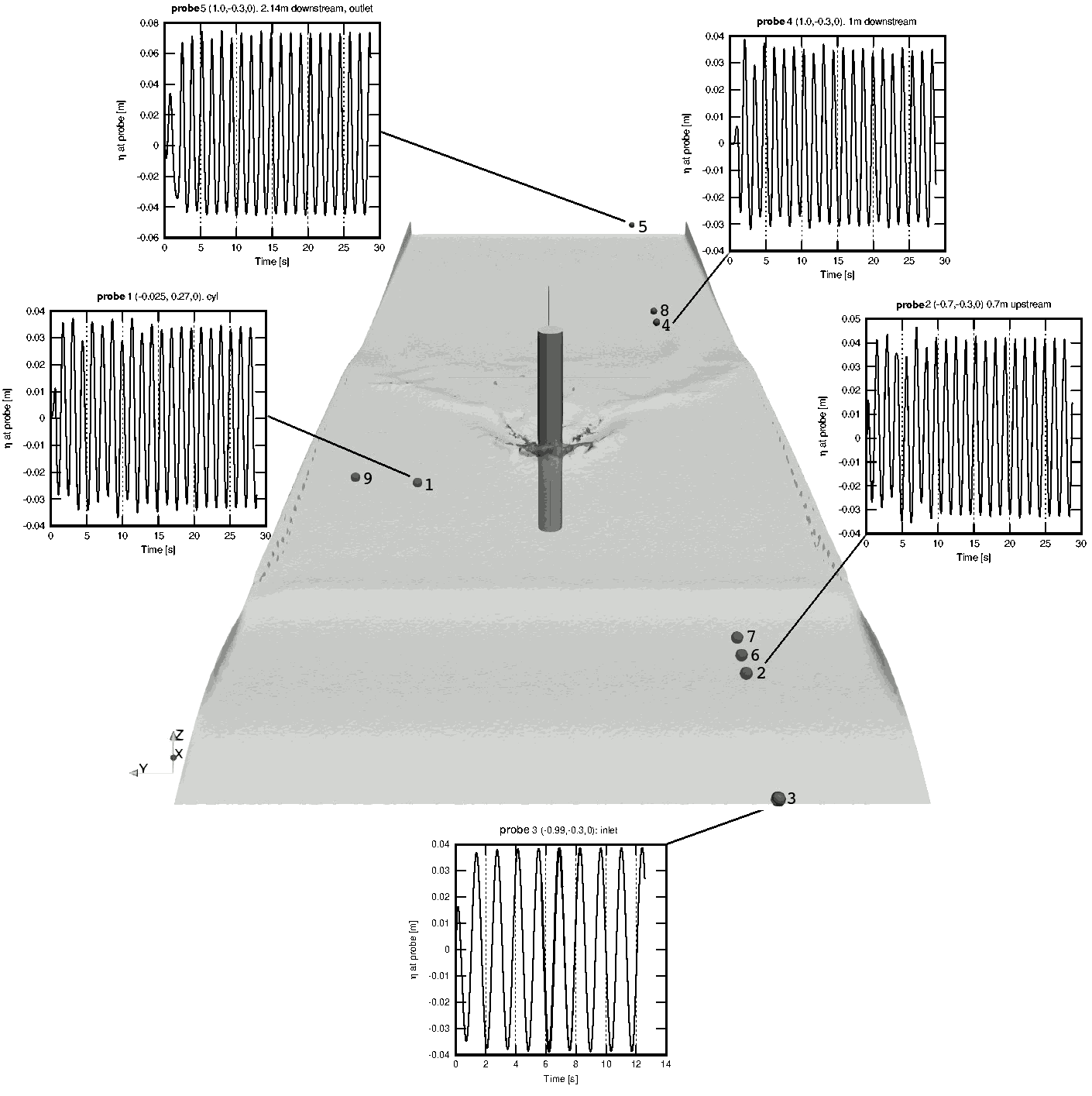}
  \caption{Positioning of point probes in the geometry of the YALES2 simulation. Each probe samples selected scalar fields every $n$-th time step ($\min(n)=1$). In this example, signals for the elevation of the $\eta$ interface are shown for four locations. In particular, the probe next to the cylinder (number 1) is visible with its elevation signal showing $2a\approx 0.05$m, very close to the specification of the experimental configuration \textit{h5c3}.. }\label{probemix}
\end{figure*}

Investigation of the water/air interface in more detail is made possible using probes.  For the current application, we have included a total of $9$ point probes, all of which are placed at the $z=0$ plane. Probes measure any scalar field chosen by the programmer; for this study we have settled upon $\eta$ and $\ub.$ Their positioning in the domain is portrayed in Figure \ref{probemix}, and the full probe list with comments is provided by the Table \ref{probetab}.

%tablework
\begin{table*}
  \begin{center}    
    \begin{tabular}{c l l l}
      Name &  Position          & Commentary            & Experiment\footnote{Coordinates given in the cylinder frame of reference} \\ \hline
      P1   &  $(-0.025,0.27,0)$ & cylinder proximity    & P1$_{exp}(-0.025,0.27,0)$\\
      P2   &  $(-0.7,-0.3,0)$   & near-inlet            & P2$_{exp}(-0.7,-0.3,0)  $\\
      P3   &  $(-0.99, -0.3,0)$ & inlet                 & P3$_{exp}(-1.5,-0.3.0)  $\\
      P4   &  $(1,-0.3,0)$      & wake waves            & \\
      P5   &  $(2.1,-0.3,0)$    & outlet                & \\
      P6   &  $(-0.65, -0.3,0)$ & pair with P2          & \\
      P7   &  $(-0.6,-0.3,0)$   & pair with P2          & \\
      P8   &  $(1.1,-0.3,0)$    & pair with P4          & \\
      P9   &  $(0.0,0.4,0)$     & ''near wall''         & \\  \hline
    \end{tabular}
    \caption{List of the point probes in the flow with their spatial coordinates and commentaries. For three of the probes, references are given to their experimental counterparts. Note that for the probe P3, the simulation probe is placed actually slightly closer to the cylinder ($1$m upstream as opposed to $1.5$m in the experiment) which is due to the simulation domain extending only $1$m upstream.}
    \label{probetab}
  \end{center}
\end{table*}

In Table \ref{probetab}, apart from specifying each probe's location, we comment briefly on its intended role; e.g. measurements of the wave elevation in the direct vicinity of the cylinder for P1. One observes that the P3 (inlet) probe is positioned in fact $1$cm downstream, which helps it convey more information about the flow physics then it would at $x=0$ which would mean a raw copy of the inlet boundary condition defined in Sect. \ref{wavegensect}. The cylinder proximity probes P1 and P9 are placed both at the $x=0$ line to coincide with the location of the solid in the channel. Probe P1 is positioned exactly in the same position as the resistive probe in the experiment, as shown in the both images of Figure \ref{manip}. The other probe, P9 is positioned in the near wall region, where it is possible to measure the elevation while avoiding the direct contact with the ripples generated at the cylinder. In other words, the probe P9 allows us to control the characteristics of the waves as seen by the solid since, before the impact, the waves are effectively two-dimensional. Probes such as P2 and P4 enable, respectively, a degree of control of the wave amplitudes upstream (but not directly at the inlet) and in the cylinder wake. In the meantime, probes P6, P7 and P8 are paired with P2 and P4 which enables etc. the calculation of the reflection coefficient.

%added 20220627
In the reference frame of the laboratory, the elevation $\eta$ seen by the cylinder probe P1 for configuration \textit{h5c3} will simply be a near-exactly sinusoidal curve with $2a=0.048$m and $T=1.43$s. This corresponds to the bottom-left curve in Figure \ref{probemix} which by comparison is characterized by slightly higher wave amplitudes  with $a\approx0.032$m. We denote that our simple wave generation method induces a small drop in amplitudes downstream, compared with that of the inlet with $a$ dropping by $10$ to $20$ percent between probes P3 and P1. This is caused by the fact that while the incoming waves are governed by (\ref{g2u}) and (\ref{g2w}), there is a possibility of backflows just downstream the inlet, which interacts with the incoming wave. We believe that the resulting wave amplitude at the structure is diminished by the said interaction, thus, we have purposefully rescaled the injected $a$ to match the experiment at the structure. Continuing the discussion of the $\eta$ evolution in the domain; one observes that signals from the probe P4 downstream resemble P1 very closely, accounting for the phase shift which is a known consequence of the structure interaction \citep{sutherland98}. Finally, we denote an amplified elevation signal at P5, suggesting that an additional treatment to damp the wave kinetic energy might be necessary should this numerical method be applied e.g. to longer domains and/or multiple structures.

\subsection{Drag Coefficient and the Pressure Force Analysis}\label{fxsect}

The pressure force exerted on the cylinder by the waves and current is calculated in the simulations by integration of the water pressure at the cylinder surface $\Gamma.$ To match the experimental measurement, we limit this computation to the $x$-component of the force, by projecting it only onto the first component of the vector $\nb_\Gamma$ normal to the boundary:

\begin{equation}\label{fx}
  F_x=-\int\limits_\Gamma \left(\nb_\Gamma\right)_x p d\Gamma.
\end{equation}

One notices also that in the examined test case, the span-wise and vertical pressure force components $F_y$ and $F_z$, while nonzero due e.g. to the separation of turbulent structures --  will not only be orders of magnitude inferior to $F_x,$ but also occasionally change sign depending on the actual wave phase. Concerning the implementation, since the unstructured YALES2 code uses tetrahedral meshing and boundary fitting, $\Gamma$ is merely a subset of the general boundary object  and  no approximations are needed to ascertain its position, unlike in certain structural mesh-based solutions (see e.g. \citealp{chenzang17}).

\begin{figure*}[ht]
  \centering
  \includegraphics[width=.9\textwidth]{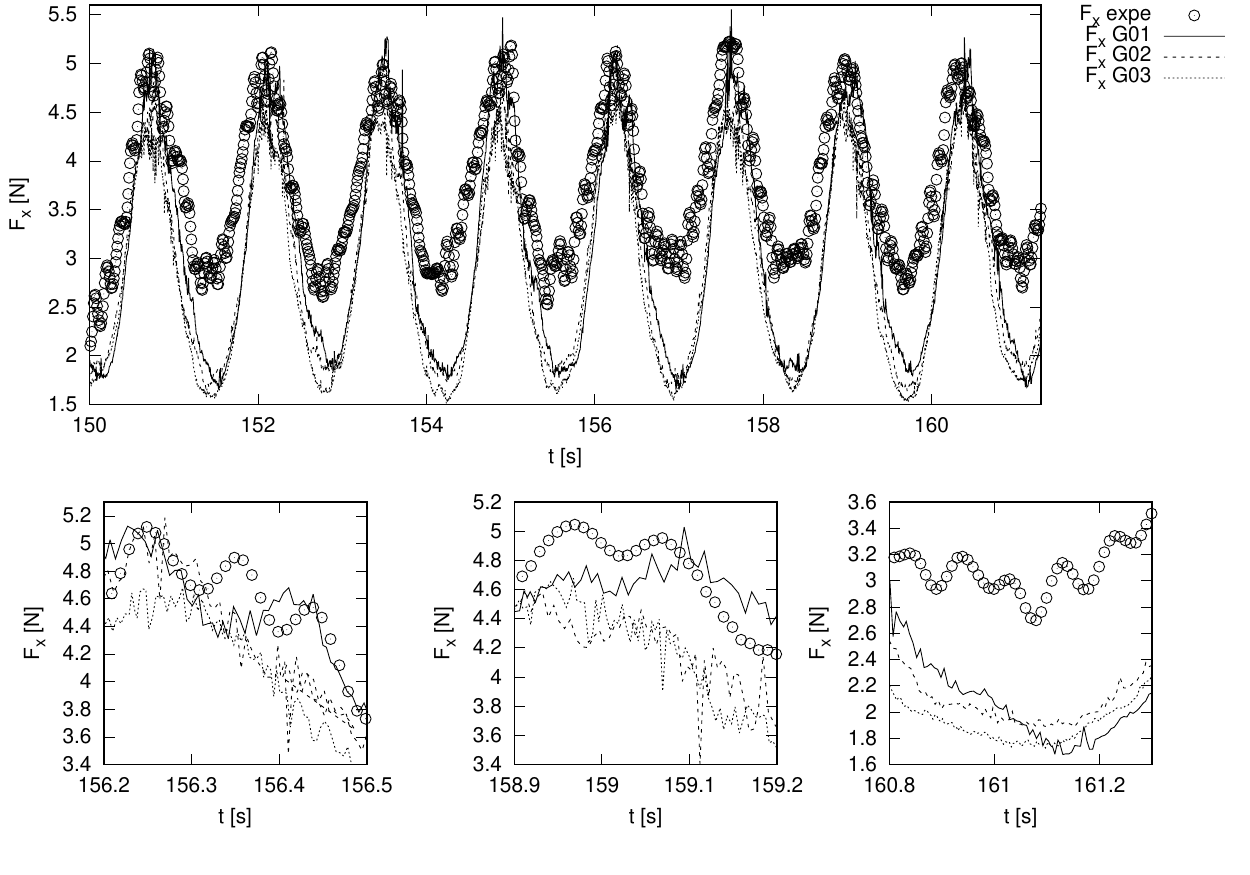}
  \caption{The pressure force $F_x$ integrated on the surface of the partially immersed cylinder. Data from the experimental measurement (circles) and numerical simulations: grids G01 (line), G02 (dashed) and G03 (dotted). Note that the time values ($t>150$s) correspond to the physical time of the carriage traction in the experiment. The insets present, from left to right, two signal peaks around $156.3$s and $159$s, respectively, and a minimum around $161.1$s.}\label{compare5}
\end{figure*}

A typical evolution of the pressure force during $11$s of the simulated time has been compared with the experimental measurement in Figure \ref{compare5}. The experimental values come from a stage at which the measurement cart has completed its acceleration, and the steady traction with the cylinder impacting subsequent waves took place. Three numerical signals are visible in Fig. \ref{compare5}, obtained using grids G01, G02 and G03. A very good correspondence is observed on all grids, as far as the signal peaks are concerned. We observe a slight disparity concerning the signal minima, as the  experimental signal has its minima at $3$N, which is over-estimated by the simulations at around $2$N. To explain this slight disparity, we note first that $F_x$ is controlled mostly by the height $h_i$ of the cylinder immersed in the water. It could therefore be inferred that the simulated waves have a minimally lower through (by which we mean $\min(\eta)$) than those in the experiment.

Figure \ref{compare5} contains results obtained using grids G01, G02 and G03 as specified in Table \ref{meshtab}. In the upper plot presenting the full signal span, we observe that both peaks and the minima decrease monotonically  in value  from G01 to G03. However, using the inset images (bottom image row in Fig. \ref{compare5}) we find that the $F_x$ grid dependency is slightly more complicated when examined in detail. Namely, around the peak at $156.25$s (image bottom left), we observe that the maximum peak value is obtained with G02 ($\min(\Delta x)=4$mm) and not G01. Further on, around the peak at $159$s, where G01 yields the highest $F_x$ values, the G03 grid ($\min(\Delta x)=3$mm) offers a better prediction than G02. Finally, the signal minimum at $161.1$s is best approximated with G02, G03 and G01 in that order. 

\begin{figure}[ht]
  \centering
  \includegraphics[width=\columnwidth]{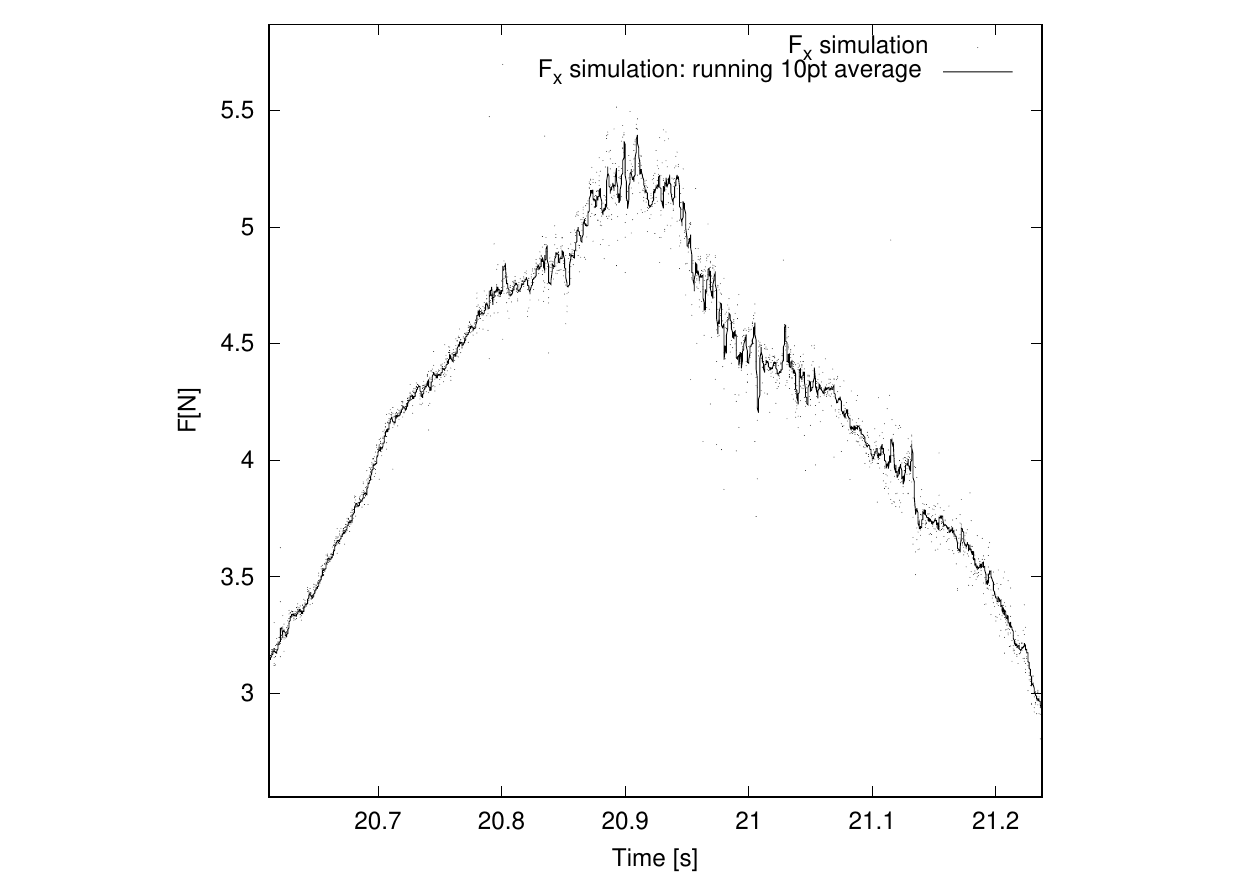}
  \caption{The pressure force $F_x$ signal obtained in the G01 simulation: raw signal (small dots) and the $10$-point walking average (line).}\label{fxnoise}
\end{figure}

The $F_x$ signal has a rather noisy character, as evidenced in Figure \ref{fxnoise}. In the Figure, we have presented, for the G01 grid,  the ten-point walking average of the force signal (continuous line) along with the raw signal (dots). It is clear that the variation in the raw signal is significant from iteration to iteration. This may suggest, naturally, that the pressure distribution on the immersed solid varies in a similar fashion; indeed the variations of the order of $0.1-0.5$N seem consistent with the turbulence levels in the boundary layer at the cylinder.

To conclude the discussion, we remark that the experimental $F_x$ signal as shown in Figure \ref{compare5} -- especially the bottom-right inset -- clearly has an additional oscillation mode of approximately $8$Hz not seen in the numerical results. After careful consideration and having re-run the simulations to no avail, we have focused our attention on the experimental measurement system to trace the source of this signal. Having performed a series of dry runs (only cylinder cart traction without the liquid) and \textit{ping-tests}, we have indeed traced a relatively weak cylinder resonance at a frequency of  $7.8$Hz (see also \citealp{valentinPHD}). This signal is superimposed onto the $F_x$ measurement visible in Figure \ref{compare5} and cannot be filtered out in post-processing (weak $7.5$Hz and $8$Hz peaks are also visible in the FFT of the experimental $F_x$ presented below in Figure \ref{spectres_exp}). Other characteristic of the experimental $F_x$ measurement is the application of a low-pass filtering algorithm (cutoff frequency of $30$Hz) which in turn removes part of the signal noise. For the future research, we remark that the solver allows inclusion of the cylinder deformation with adaptive meshing which could help resolve the mode present in the experiment.

\subsection{Vorticity evolution}

\begin{figure}[ht]
  \centering
  \includegraphics[width=1.0\columnwidth]{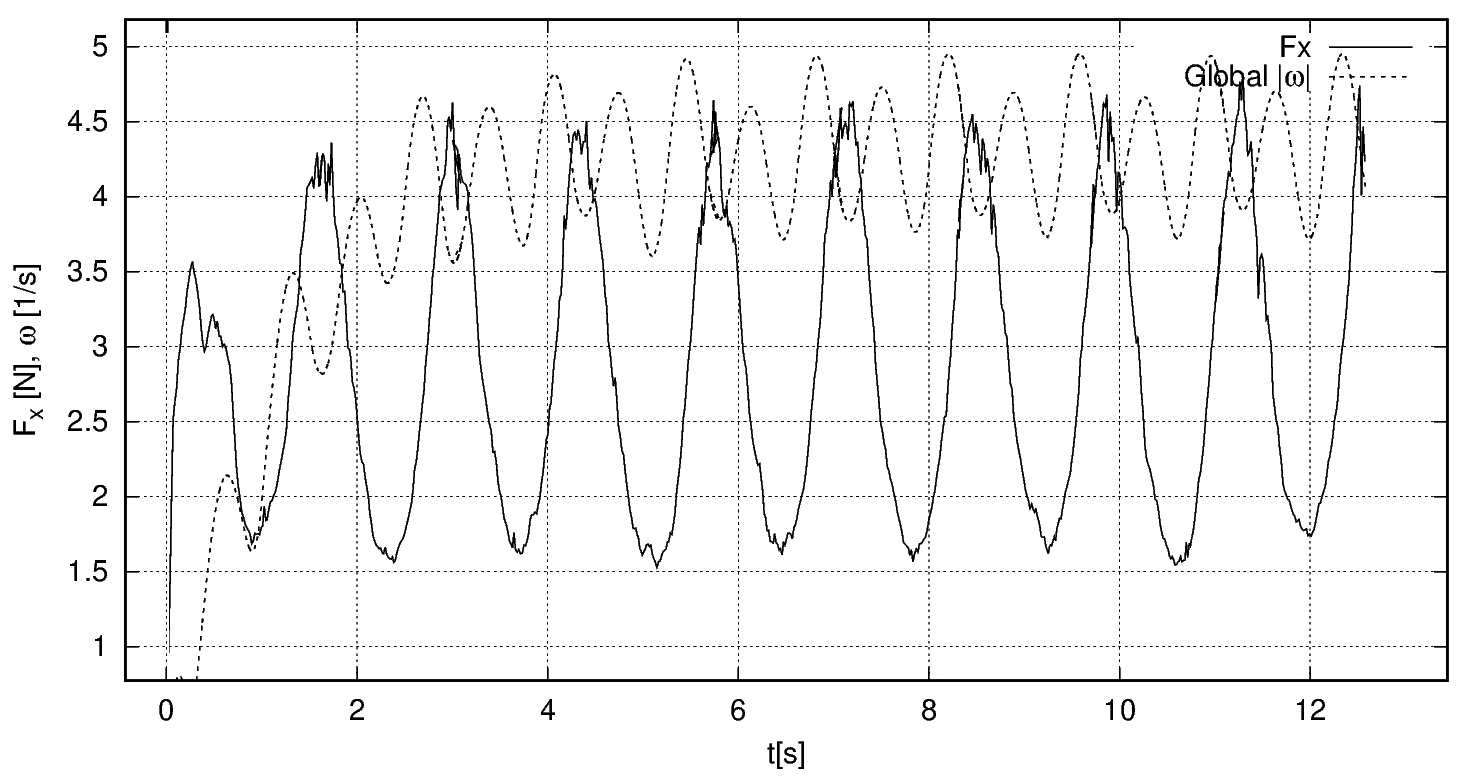}
  \caption{Superimposed curves of the $F_x$ and the vorticity integral obtained in the G03 simulation. The vertical axis unit is Newtons, however it so happens that $\int\omega dV$ oscillates around $4\cdot1/s$, meaning that no scaling is used in the plot and the $y$ tics are accurate also for the vorticity unit. }\label{vortfx}
\end{figure}

The temporal evolution of the cylinder wake vorticity is presented in Figure \ref{vortfx} (dashed line). It should be precised that in the numerical simulation results we have found that the temporal evolution of global vorticity in the \textit{whole} domain is indistinguishable from vorticity integrals calculated over wake sub-domain. In other words, the bulk vorticity is contained in the wake, thus we are presenting the global $\omega$ integral in Figure \ref{vortfx}. In order to be able to trace the interaction between the major waves, the cylinder and the wake vorticity, we have superimposed onto the plot the $F_x$ curve obtained from the G03 simulation (dotted line in Fig. \ref{compare5}). As can be readily observed in Fig. \ref{vortfx}, there is a strict synchronisation between the $\omega$ maxima and that of the pressure force. Namely, two $\omega$ peaks are found for each wave period. Thus, vorticity minima are observed whenever wave crests or throughs impact the cylinder, with $\omega$ peaking whenever a wave slope is directly upstream or downstream. This behaviour has previously been described e.g. by Arabi et al. \citep{Arabi2019} concerning an  interaction of waves with a  cubical obstacle which they studied numerically. Upon the impact of the incident wave at the windward cylinder side, vortical structures are formed there due to the creation of the water elevation (damming). Also, at this stage the structures near the cylinder bottom intensify. This leads to the creation of the first $\omega$ peak, while the second peak forms after each incident wave's crest passes the cylinder, leading to the intensified creation of the detached vortical structures on the leeward side. The vorticity evolution is affected also by the reshaping of the cavity (change in the wake angle $\alpha$, see Figure \ref{halfangels}).

In context of the vortical structures detaching from the cylinder, we denote that the velocity gradient in the boundary layers has a certain grid dependency. This poses a challenge on homogeneous tetrahedral grids (see \citealp{mukha21}). In the case of the G0-G03 grids used here, we have denoted slow convergence of the wall-normal velocity gradient with the values of $Y+\le 20$ near all boundaries (notably, $Y+<20$ next to the $z-$ wall (flume bottom) as well as cylinder surface). However, both  purely geometrical aspects (such as the grid skewness) as well as flow physics (boundary layer separation, cusps in geometry), may give rise to occasionally higher $Y+$ values. We have denoted the $\max(Y+)\le 90$ for all G01, G02 and G03 grids, with the maximum values corresponding either to highly skewed tetrahedra, or the cusps e.g. at the bottom of the immersed cylinder. 
 
\subsection{Spectral Analysis}\label{spectralsect}

In this section, we briefly present the results of the investigations using spectral information obtained mainly by transforming the probe signal. In two-phase flows, while the Fourier transforms (FFT, DFT) of the signals sampled \textit{far from the interface} are a well accepted tool \citep{pope}, the number of works employing it to actually study the interface itself is far inferior. Ling et al. \citep{ling15} have used it on interface height samples obtained from a shearing flow. Our application is similar - for both cases the key fact is that the interface is initially flat, so that the height (or distance) - measuring probe can yield meaningful information on interface wrinkles, oscillations or deformations as it always oscillates around the initial zero-height position. In applications such as atomization \citep{ling15} or even Rayleigh-Taylor instability's later stages where the bulk of the liquid undergoes a complete displacement \citep{aniszewski2014caf} this kind of application would not be possible.

\begin{figure}[ht]
  \centering
  \includegraphics[width=.55\textwidth]{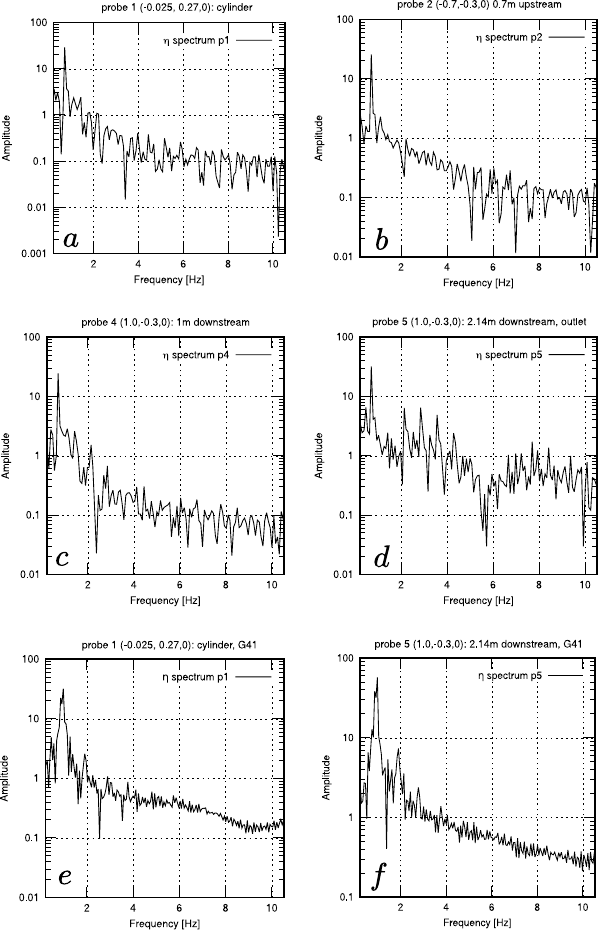}
  \caption{Spectra of the $\eta$ signal for six of the probes using G03 (a-d) and G41 (e-f) grids.}\label{elevspect}
\end{figure}

That said, Figure \ref{elevspect} presents the discrete FT on six elevation $\eta$ signals obtained by the probes throughout the simulation. The spectra only supply information for $f<10$Hz, as above that frequency, for this flow configurations, no signals of amplitude stronger than noise are expected to appear in the flow. In Figure \ref{elevspect}, the spectra of the G03 simulation (Fig.\ref{elevspect}a-d) and the G41 grid (no cylinder, Fig.\ref{elevspect}e-f) are visible. All spectra are characterized by the dominant first harmonic ($T=1.43$s or $f=0.69$Hz, the generator frequency). Unlike upstream (Fig.\ref{elevspect}b), all spectra downstream the cylinder starting with the structure probe itself (Fig.\ref{elevspect}a,c,d) contain the second harmonic near $2f=1.38$Hz. We could associate this frequency with the peaks of $\omega$ as described in the context of Figure \ref{vortfx}. The third harmonic ($3f=2.07$Hz) is well visible further downstream, suggesting it may originate e.g. from the action of the evolving wake edges. Finally, the outflow probe spectrum (Fig. \ref{elevspect}d) has evidence of further harmonics. In contrast, the flow for which there was no fluid-structure interaction (Fig. \ref{elevspect}e-f) save for the containment effect of the channel \citep{potg22}, is characterized by lower spectral content limited to the third ($1$m downstream, Fig.\ref{elevspect}e) and/or fourth (outlet) harmonics. Higher frequency content is largely absent in this flow.

\begin{figure}
\includegraphics[width=1.0\columnwidth]{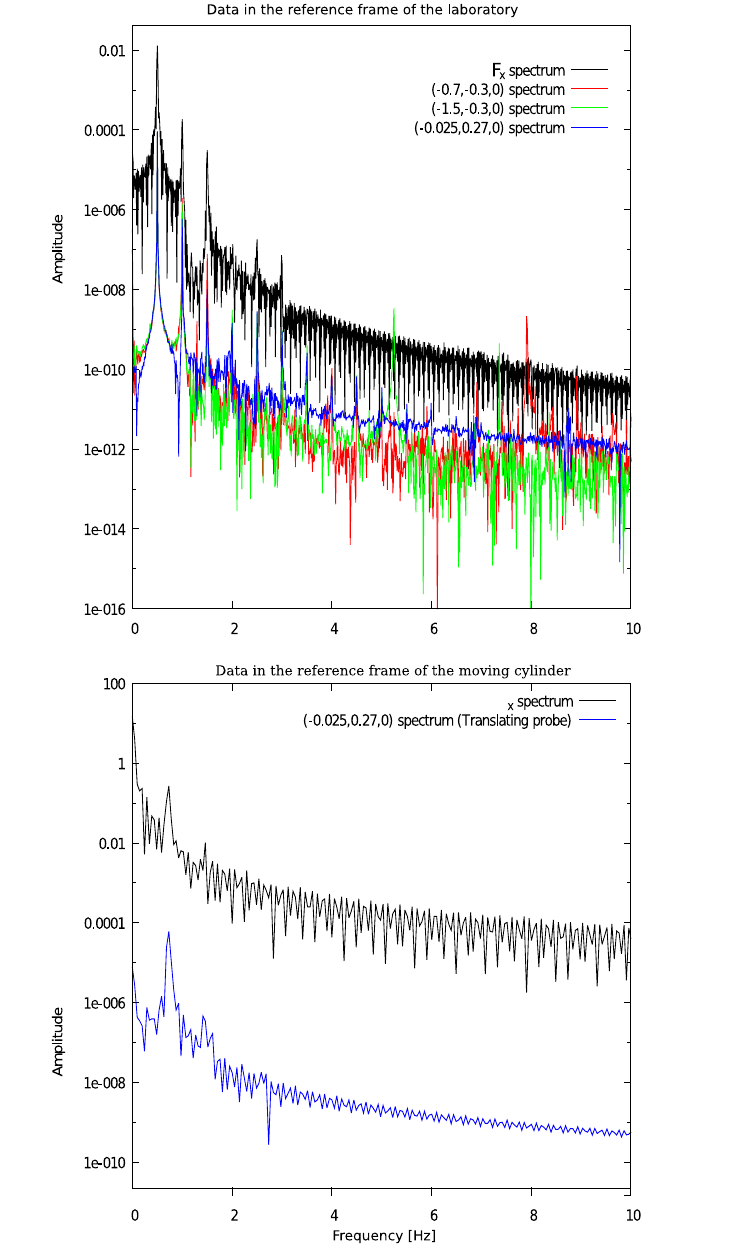}
 \caption{Top: Elevation $\eta$ spectrum of the P1$_{exp}$-P3$_{exp}$ probes and drag force $F_x$ (black line). Bottom: probe P1$_{exp}$ elevation (blue)  and drag force spectra (black) for the cylinder translating at the velocity $u_0 = 0.8375 $ m/s}
 \label{spectres_exp}
\end{figure}

The experimental spectra of the incoming waves (in the frame reference attached to the laboratory), the drag force and the probe P1 attached to the moving cylinder are depicted in Figure \ref{spectres_exp}. In comparing the experimental signal spectra presented in Fig.\ref{spectres_exp} with Fig.\ref{elevspect}, we use the blue curve presented in the upper picture of Fig.\ref{spectres_exp}, i.e. the $\eta$ signal associated with the P1$_{exp}$ probe. This curve corresponds directly to Fig. \ref{elevspect}a. Taking into account the difference in the amplitude normalization, we notice a similarity in the slope of both spectra. Also, the second harmonic visible in Fig.\ref{elevspect}a has a far lower amplitude compared to Fig.\ref{spectres_exp}. Obviously, the experimental data contains at least ten clearly visible harmonic peaks, and has excellent peak to noise ratio, the lack of which presents an obstacle in interpreting Fig.\ref{elevspect}a.

\subsection{Analysis of Wave Reflection}\label{reflectsect}

Both numerical and experimental data in our study allow for estimations of the characteristics of the reflection coefficient. In the simulation, probe elevation signals are used, allowing for a free choice of the involved probe pairs or triplets. Subsequently, we employ the methodology of \cite{goda1976} to resolve the composite waves into those incident and reflected. The calculation is performed in the spectral space, with the restriction that the elevation records must originate from probes spaced closer than $\lambda/2.$  For the experimental results measured using the resistive probes \citep{mansard1980}, the exact same methodology to calculate $C_r$ is used.

Thus, for the experimental data in the \textit{h5c3} case, the reflection coefficient of the incoming waves is estimated at $C_r \simeq 0.0451$ using probes P2$_{exp}$ and P3$_{exp}.$ In the simulations, the base grid G01 yields $C_r\simeq 0.35$ using probes P2 and P6. However, using the G03 with the same probes involved, we have obtained $C_r\simeq 0.21,$ which is still roughly $5$ times the experimental value. It seems an expected result considering the simplicity of our wave generation scheme and the lack of more rigorous outflow damping. It has to denoted that the probe signal is -- both for the experimental and numerical data -- truncated to remove first seconds of the flow corresponding to the cart acceleration stage\footnote{Certain sensitivity to the exact choice of the truncation time takes place, thus we report averaged $C_r$ values here.}.

Additionally, we have  performed studies using the grids G41 and G42. Results using  the G41 grid indicate that $C_r$ actually grew to $0.5$, confirming that the reflection is produced at the outlet in all tested simulations (and not only the default configurations with the cylinder G01-G03). Since the G01 simulation contains the cylinder and G41 does not, and both the measurement points lie upstream the cylinder, we conclude that the reflected wave was partly dispersed by the cylinder in G01-G03 before reaching the location of the probes. This explains the growth of $C_r$ using the G41. Additional simulations were performed using the G42 grid (see Table \ref{meshtab}) which contains a longer flume and applies a 1m long sponge layer, i.e. a coarsen grid with larger tetrahedra. Unfortunately, the results for this numerical setup yield $C_r>1$; this amplification can be explained twofold: first, the sponge layer is insufficient to damp the outgoing waves using only the phenomenon of numerical viscosity (i.e. numerically induced $E_k$ dissipation). Second, the reflected wave moves back upstream in the G42, the action no more that efficiently dispersed by the obstacle as in the case of G01-G03.

Thus, we conclude that obtaining an exact match of the $C_r$ with the experimental data could be obtained either by additional damping techniques near the outflow (e.g. porous ramps as in the laboratory setup), or the further increase of the spatial resolution, in hope that the dependency of $C_r$ on the mesh size -- as observed progressing from G01 to G03 -- continues to the finer meshes.

\subsection{Statistical Analysis}\label{statsect}

Instantaneous results limit our possibilities in discerning between transient and periodic phenomena especially at the small scales. Thus, e.g. to further analyze the possible presence of reflected components in the composite waves present in the domain, we have employed statistical techniques such as phase averaging. The technique is applicable to periodic processes (see \citealp{wernert} or for a similar application \citealp{carvaro}) and utilizes temporally averaged data grouped in bins spaced by $T$ (in which case $T$ is the inherent process period, which in our case is naturally the wave period). More specifically for the time-varying signal such as the elevation $\eta(t)$, the phase average can be estimated as

\begin{equation}\label{phasavdef}
  \langle \eta(t,N)\rangle_{est} = \frac{1}{N}\sum\limits^N_{i=1}\eta (t+(i-1)T),
  \end{equation}

where $N$ is the number of the averaged periods which should be as large as possible. As explained by \cite{wernert}, the desirable $N$ values are of order of hundreds for the experimental signals. However, in this study we are limited in terms of the CPU cost, limiting the temporal span of a typical G01 simulation  to $t_{end}\in\lbrack 13T,30T\rbrack$ with $t_{end}$ being even smaller for the more costly grids such as G03. Thus, results presented below should be treated as rough approximations in terms of the convergence of (\ref{phasavdef}) to $\langle \eta(t,N)\rangle.$

\begin{figure*}[ht]
  \centering
  \includegraphics[width=1.0\textwidth]{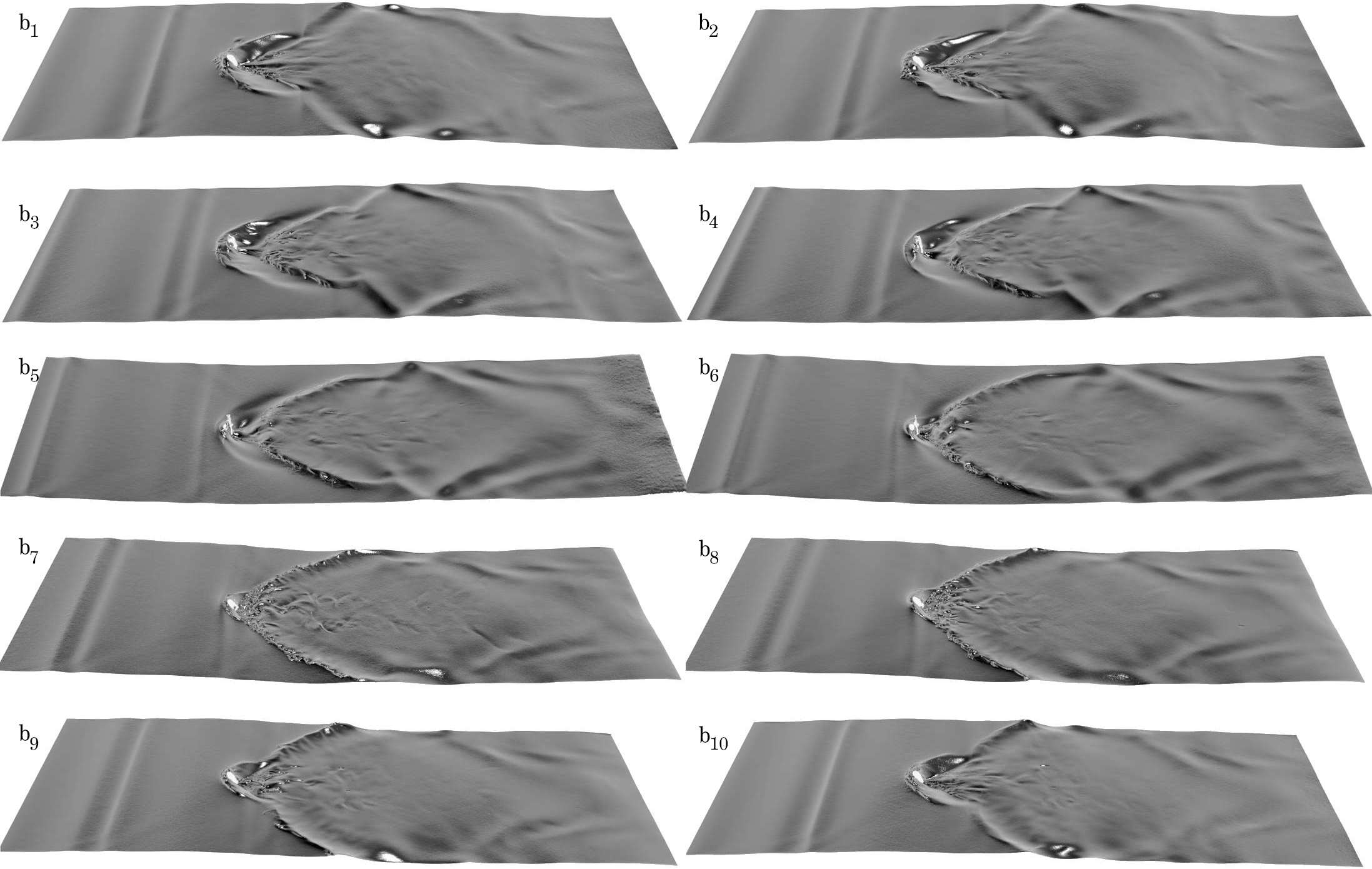}
  \caption{The phase-averaged mean values of the volume fraction function $\chi(\phi)$ for the simulation on the G03 grid. (The $\chi(\phi)=0.5$ isosurfaces are shown.) Each picture corresponds to a $T/10$-sized collection bin, where $T=1.43$ (e.g. the top left image is obtained using data from the bin $b_1$ where $t\in \lb 0+nT, 0.143+nT\rb$). The data has been collected over the entire length of the G03 simulation (approx. $12$s of physical time, thus $n=1\dots8$).}\label{phiavg3}
\end{figure*}

An example of the phase-averaged mean estimation of the elevation is presented visually in Figure \ref{phiavg3}, which showcases the isosurfaces of the $\langle \phi \rangle_{est}$ for $N=8$ obtained in the G03 simulation. Since the images in Fig. \ref{phiavg3} correspond to all ten bins $b_1\dots b_{10}$ spanning the entire period $T,$ we observe how the $\lambda\approx 2.34$m wave\footnote{The wavelength is given in the frame of reference associated with the cylinder. In the laboratory frame of reference $\lambda\approx 3.69$.}  passes through the cylinder proximity in the $b_1$ image, leaves the simulation domain in the $b_5$ and $b_6$ bin images to reenter by the $b_7.$ Other notable features include fluctuations in the cylinder trace (\textit{v-shape}) angle, which we will discuss below.

As made clear by Figure \ref{phiavg3}, the numerical setup results in a low-amplitude, ``w-shaped'' reflection wave close to the outlet which remains nearly static during the  flow period $T.$ This is consistent with the relatively high values of the $C_r$ as described in Sect. \ref{reflectsect}; we attribute this to the rather short computational domain, and the lack of numerical wave suppression towards the channel outlet. However, we were not able to confirm the possible propagation of the reflected wave towards the immersed cylinder.  Of similar (numerical) origins are the two-dimensional, stripe-shaped disturbances visible upstream of the cylinder in Figure \ref{phiavg3}. These disurbances result from the small-scale disturbances at the domain inlet, and are more prominent when $a$ (the injected wave amplitude) is higher. In most cases, however, as visible in Figure \ref{phiavg3} (see e.g. $b_6$) these formations dissipate before impacting the structure, and their effect on data such as the $F_x$ spectra presented in Sect. \ref{spectralsect} is negligible.

\begin{figure}[ht]
  \centering
  \includegraphics[width=0.9\columnwidth]{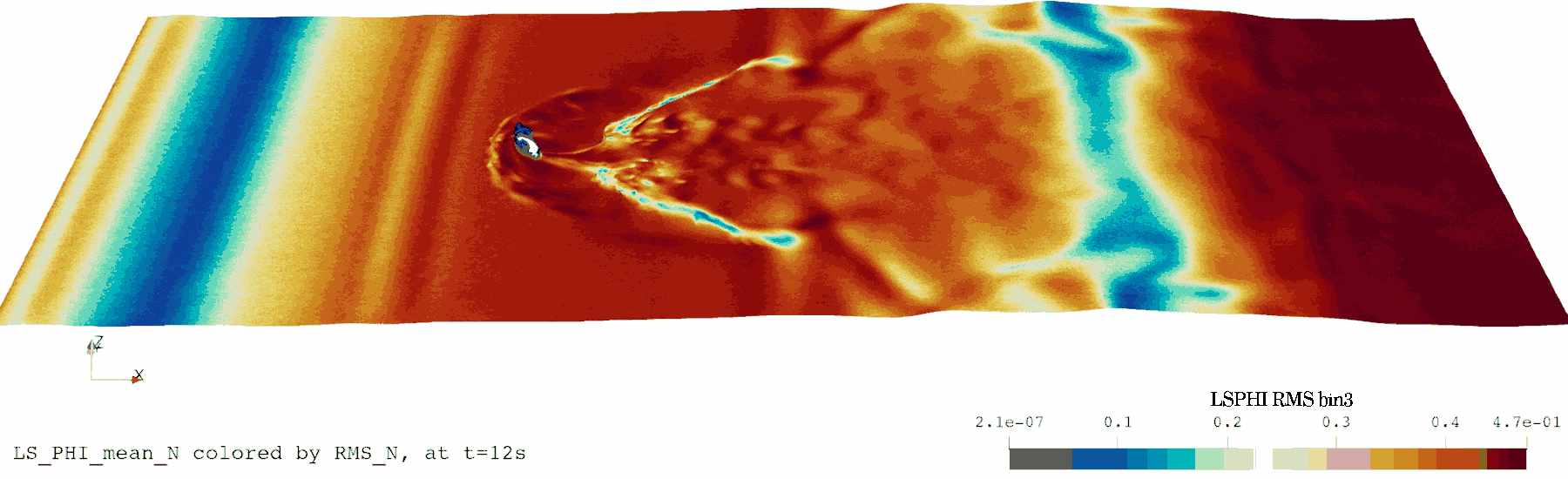}
  \caption{The phase-averaged mean value of $\chi(\phi)$ for the G03 simulation. Image of the $\chi=0.5$ iso-surface presented previously in Figure \ref{phiavg3}, with the color scheme corresponding to the phase-averaged standard deviation of $\phi$.}\label{phiavg4}
\end{figure}

The isosurface corresponding to the third bin $b_3$ of the phase average (Figure \ref{phiavg3}) is reprinted in Figure \ref{phiavg4} with a different color scheme. This time, the color corresponds to the root mean square $\langle \phi^2\rangle^{1/2}$ of $\phi$ obtained with phase averaging\footnote{Referred to also as the standard deviation, i.e. the square root of the variance, see \cite{pope}, p.41.}. Using such color scheme allows us to focus our attention on the deviation of the interface position from its phase-averaged mean position for the chosen bins (i.e. selected intervals of the wave period). Figure \ref{phiavg4} confirms, in this way, that the areas with lowest period-to-period variance correlate with the crests and throughs of the $\lambda=2.34$m waves. Interestingly, an area of the v-shape in the downstream trace is also coloured blue in Fig. \ref{phiavg4} suggesting that the v-shape evolution is strictly periodic and not subject e.g. to turbulence in the cylinder wake at least at the tested Reynolds and Froude numbers.  It is also interesting to note that the near-outflow corresponds to maxima of $\langle \phi^2\rangle^{1/2}$, i.e. the variance of the interface position is high near the outflow during the simulation.

\begin{figure}[ht]
  \centering
  \includegraphics[width=0.9\columnwidth]{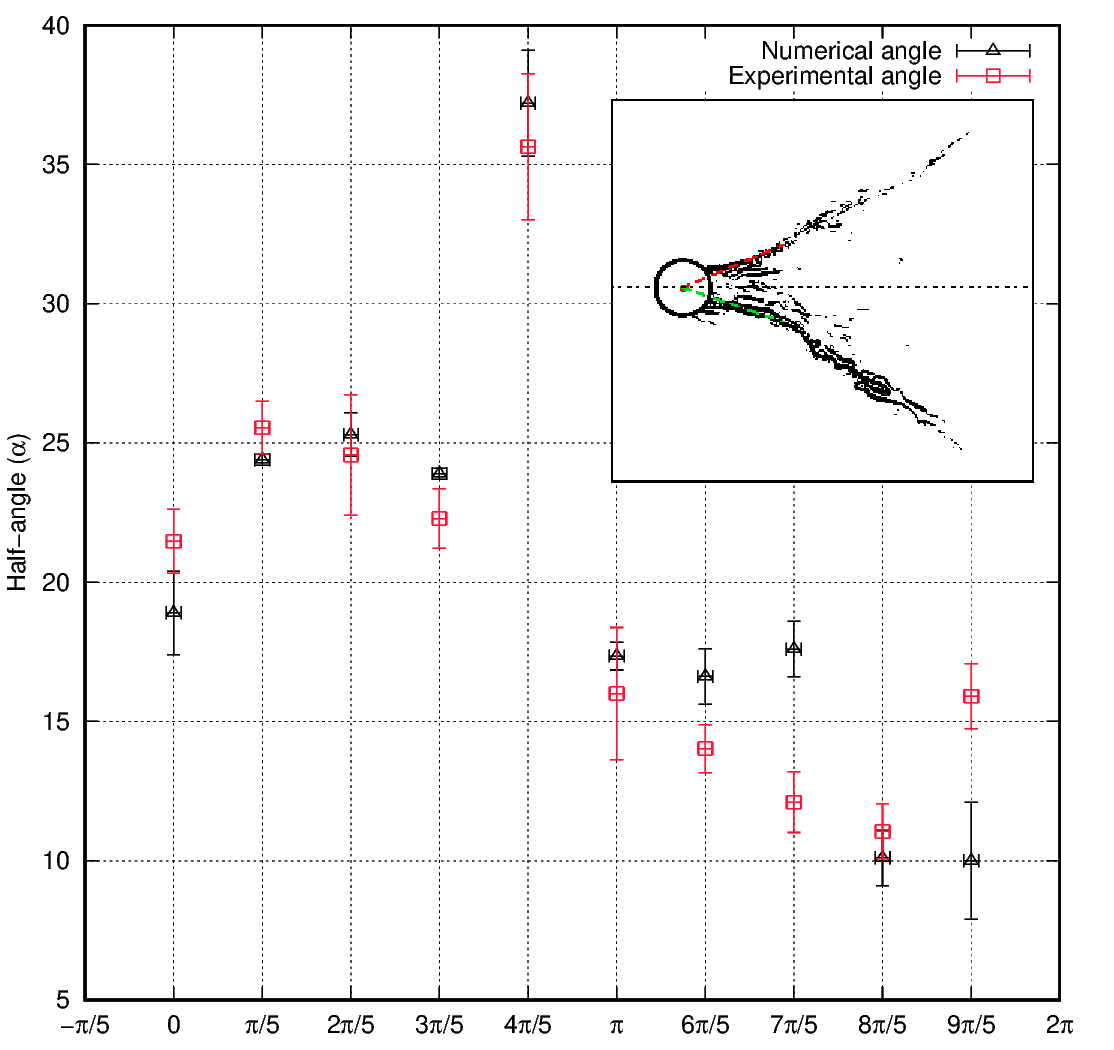}
  \caption{The ''v-shape'' half-angle study: experimental result (red squares) and simulation (black triangles). The inset presents the angle-measurement technique as explained in the text.}\label{halfangels}
\end{figure}

In order to further validate the numerical result using the phase averaging, we finalize the section with a comparison between the numerical and experimental results concerning the half-angle of the "v-shape", which is the geometrical trace created by the liquid folds behind the cylinder. The comparison is presented in Figure \ref{halfangels}. The comparison is performed as follows. For each of the bins, the $\langle \chi\rangle=0.5$ iso-surface, representing the mean interface position, is examined. Using the top view (i.e. along the $z$ axis), we first process the image with edge detection algorithms and subsequently measure the angle between the domain $x$ axis and the liquid folds (measurement is taken $2D$ downstream). The procedure is presented symbolically in the inset of Fig. \ref{halfangels}: we can see two diagonals marking the liquid folds. (In this particular example (bin $10$), the folds are not strictly symmetrical, in which case an average value is presented in Fig. \ref{halfangels}.) The error bars in the numerical case are obtained using the standard deviation information for the $\chi(\phi)$ function.

Analyzing the Figure \ref{halfangels}, we observe a relatively close fit between the half-angle values for $7$ of $10$ bins. Notably, the wide ''v-shape'' half-angle $\alpha>35$ for the phase $4\pi/5$ is well captured by the simulation. This moment in the phase-average corresponds to the onset of the impact of the $\lambda=2.34$m wave on the cylinder, the actual wave crest passing through the structure takes place for the next measurement point ($\pi$) which is also well captured. The points $6\pi/5$ and $7\pi/5$, corresponding to the moments in the flow where the large wave crest has moved downstream the cylinder,  exhibit discrepancies between simulation and experiment. As shown in the vorticity analysis (see Figure \ref{vortfx}) once the crest clears the cylinder a vorticity peak appears, suggesting a maximum vorticity production in the wake. Since this mechanism is severely under-resolved in our simulation, we suppose that the vorticity production mechanism underwater may imprint on the surface eddies in the wake influencing the half-angles. Thus, the discrepancies in Fig.\ref{halfangels} could be of numerical origin and due to lack of resolution. That said, the general $\alpha$ evolution seems well predicted by the simulation within these limitations, as evidenced by Fig. \ref{halfangels}.

The work of \cite{rabaudmoisy} proposes a method to analytically approximate the (half) wake angles based on the Froude number. They show that for $\Fr>\Fr_c\approx 0.49,$ the angle expresses as:

\begin{equation}\label{rabaud}
  \alpha=\tan^{-1}\frac{\sqrt{2\pi\Fr^2-1}}{4\pi \Fr^2-1},
\end{equation}

which is applicable to our configuration since $\Fr=1.2$, leading to the ''average'' wake angle of $\alpha=18\degree.$ Applying (\ref{rabaud}) to \textit{instantaneous} velocities probed in the interface vicinity with the $P_1$ probes, we denote, that $u\in\lbrack 0.4,1.6\rbrack$ there resulting in ''local'' values of the Froude number between $0.57$ and $2.28$, hence $\alpha_{min}\approx 14\degree$ and $\alpha_{max}\approx 36\degree,$ which is in good comparison with Figure \ref{halfangels}.

\begin{figure*}[ht]
  \centering
  \includegraphics[width=0.9\textwidth]{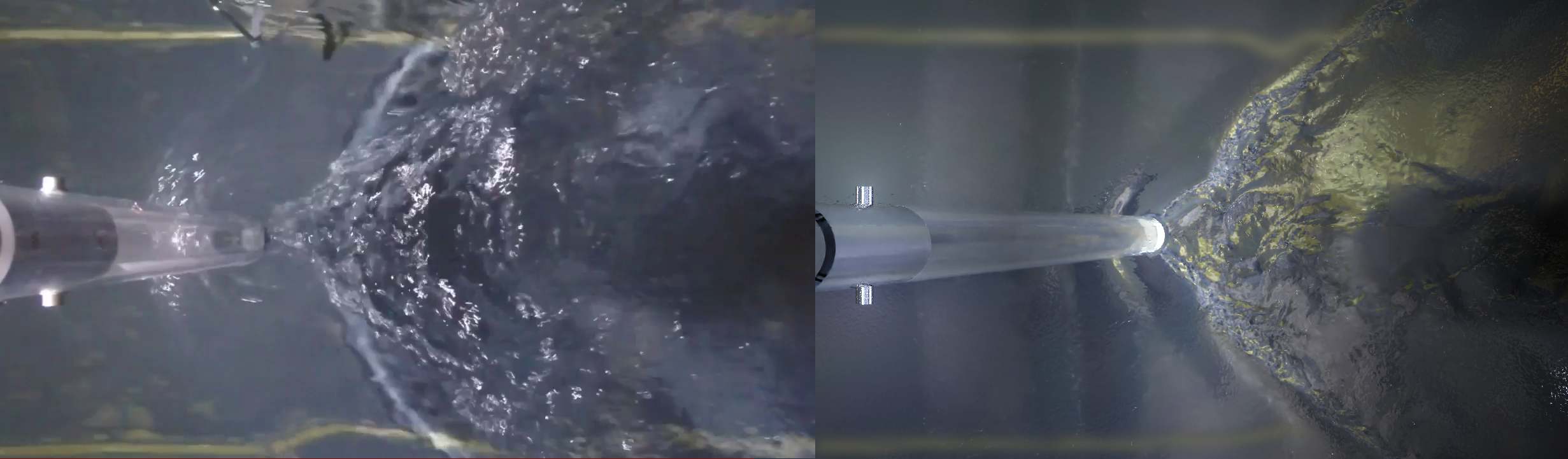}
  \caption{A visual comparison between instantaneous results of the experimental (left) and numerical (ritght) study. The image corresponds to $t\approx 152.6$s in the laboratory frame of reference.}\label{render}
\end{figure*}

Final comparison between the numerical and experimental results is found in Figure \ref{render}, showcasing a rendered view of the simulation (right) with the top-view photography obtained during the experimental acquisition (left). The laboratory view has been obtained with a portable camera fitted directly to the cylinder mount, with a similar (although not identical\footnote{This is evident from the fact that far more liquid is visible upstream the cylinder in the rendered image.}) position used for the rendering. The view corresponds to an instantaneous interface geometry at $t\approx 152.6$s (in the laboratory frame of reference), and a wider v-shape angle. In the photograph visible in Fig. \ref{render}, the horizontal yellow-green lines correspond to the physical flume walls, while in the rendering they are artificially superimposed (the actual camera angle did not include domain walls at the $y$ extremities). Concerning the interface characteristics, while the v-shape is comparable in angle, there seems to be more of a structure visible in the wake $4-5$D downstream in the experimental image. This may suggest that the sub-surface wake eddies are similarly weaker in the numerical result, as noted in the context of Fig. \ref{halfangels}. The upstream liquid elevation appears in both images; also a similar  interface structure (small-length waves) is found in both images further downstream.
 
\section{Conclusions}
%\lipsum[12]

In this paper, we have presented a numerical tool for simulating the interactions of the waves with a submerged structure. Simulations involve a partially submerged cylinder towed in a flume containing wavy flow. Our study involves an experiment wherein that exact setup is used in a full length flume. The experimental campaign conducted at the LOMC laboratory includes measures of multiple cylinder radii at varying $\Re$ and $Fr$ values; in our paper the focus has been on a single configuration where $D=0.05$m and $\Re\approx 41875.$ The simulation domain contains only a section of the flume, and the waves are generated at the inlet. 

We have observed a very good simulation-to-experiment comparison concerning wave geometry, macroscopic flow statistics and, most importantly, the hydrodynamic load onto the cylinder expressed in terms of the pressure force $F_x.$ From a numerical point of view, the faithfulness of the $F_x$ calculation seems to rely mostly on the correct immersion depth for the cylinder, as well as the mesh quality. This yields the proper integration surface on the solid boundary and improves the match with the experimental data. This, subsequently, means that factors such as fitted boundary smoothness (lack of any artificial roughness) and/or the quality of the boundary tetrahedra may be equally influential as the minimum grid size $min(\Delta x).$ 

We have found that the vorticity generation mechanism is closely linked with the position of the incident waves, which is consistent e.g. with previous studies \citep{valentinPRF,valentinFLUIDS}. Each passing wave has two associated $\omega$ peaks in the cylinder wake zone as described by \cite{Arabi2019} -- one at the moment of the impingement, and another one associated with the vortical structure separation and the growth of the ''v-shape'' angle $\alpha$ -- after it passes the solid structure. Our study can be considered highly resolved in comparison to certain previous works \citep{carvaro}.

Using the information from the probe points placed in the calculation domain, we were able to establish a close correspondence with the experimental results in the spectral space. A singular, not-matching  peak was found in the experimental data caused by the resonance of the cylinder traction cart. Data from the flow probes has allowed to find a closely matched $\eta$ evolution, as well as control phenomena such as wave reflections. Due to significant values of the reflection coefficient $C_r$ measured by the probes close to the inlet, we must conclude that using the $3.14$m-long domain results in significant levels of reflection. A 'w-shaped' reflected wave is visible in the phase-averaged geometry plots, and  its removal should be the natural task for the continued research. In fact, during the work described in this paper, longer computational domains (up to $11$m) have been tested, including variants with outflow ramps. However, this solution was deemed too computationally expensive for the current campaign. Separately, wave absorption zones \citep{han15} were tested, however the latter solution did not mitigate the reflection phenomenon in preliminary tests. Thus, in these authors' opinion, a solution combining both a ramp -- with a form of simulated porosity --  and absorption zones should be tested in a separate work. The same goes for techniques such as fully dynamic AMR. 

That said, the goal of the paper was to demonstrate that even with a relatively short flume section simulated, it is possible to capture fundamental wake physics, especially if the spatial resolution is significant close to the embedded structure. Reusing the presented method can present an interesting opportunity e.g. for testing higher Reynolds numbers, which in turn would place the flow in the regime characterized e.g. by significant air entrapment in the cylinder wake. In broader terms, it allows a more wide-ranging parametric studies of the wave-structure interaction in a two-phase flow.

\section*{Acknowledgements}
The authors acknowledge the funding in the frame of the LabEx EMC3-2019-PC \textit{STRUCTIMM} project at the CNRS. Calculations were performed using the resources of the CRIANN (Centre Régional Informatique et d'Applications Numériques de Normandie) notably the \textit{Myria} supercomputer (2020-2021). More calculations were performed using the TGCC (Très Grand Centre de calcul du CEA) supercomputer \textit{Irene Joliot-Curie} (2021). 

\bibliographystyle{elsarticle-harv}
\bibliography{aniszewski}

\end{document}